%% file: SIAP_PDD_arxiv2025Dec.tex
\documentclass[hidelinks,onefignum,onetabnum]{siamart251216}

\usepackage{lipsum}
\usepackage{amsfonts}
\usepackage{graphicx}
\usepackage{epstopdf}
\usepackage{algorithmic}
\ifpdf
  \DeclareGraphicsExtensions{.eps,.pdf,.png,.jpg}
\else
  \DeclareGraphicsExtensions{.eps}
\fi

\newsiamremark{remark}{Remark}
\newsiamremark{hypothesis}{Hypothesis}
\crefname{hypothesis}{Hypothesis}{Hypotheses}
\newsiamthm{claim}{Claim}

\headers{Pointwise Distance Distributions for detecting near-duplicates}{D. Widdowson, V. Kurlin}

\title{Pointwise Distance Distributions for detecting near-duplicates in large materials databases\thanks{To appear in SIAM Journal on Applied Mathematics, doi:10.1137/25M1736657.
\funding{Royal Society APEX fellowship APX/R1/231152, New Horizons grant EP/X018474/1}}}

\author{Daniel E. Widdowson\thanks{Department of Computer Science, Liverpool, UK 
  (\email{D.E.Widdowson@liverpool.ac.uk}).}
\and Vitaliy A. Kurlin\thanks{Department of Computer Science, Liverpool, UK  (\email{vkurlin@liv.ac.uk}, \url{http://kurlin.org}).}
}

\usepackage{amsopn}

\usepackage{lineno}

\ifpdf
\hypersetup{
  pdftitle={Pointwise Distance Distributions for detecting near-duplicates},
  pdfauthor={D. Widdowson and V. Kurlin}
}
\fi

\input commands.txt

\begin{document}

\maketitle

\begin{abstract}
Many real objects are modeled as discrete sets of points, such as corners or other salient features.
For our main applications in chemistry, points represent atomic centers in a molecule or a solid material.
We study the problem of classifying discrete (finite and periodic) sets of unordered points under isometry, which is any transformation preserving distances in a metric space.
\smallskip

Experimental noise motivates the new practical requirement to make such invariants Lipschitz continuous so that perturbing every point in its $\ep$-neighborhood changes the invariant up to a constant multiple of $\ep$ in a suitable distance satisfying all metric axioms. 
Since the given points are unordered, the key challenge is to compute all invariants and metrics in a near-linear time of the input size.
\smallskip

We define the Pointwise Distance Distribution (PDD) for any discrete set and prove, in addition to the properties above, the completeness of PDD for all periodic sets in general position.   
The PDD can compare nearly 2 million crystals from the world's five largest databases within 2 hours on a modest desktop computer.
The impact is upholding data integrity in crystallography because the PDD will not allow anyone to claim a `new' material as a noisy disguise of a known crystal.
\end{abstract}

\begin{keywords}
isometry classification, complete invariant, continuous metric, periodic crystal
\end{keywords}

\begin{MSCcodes}
74E15, 68U05, 51N20
\end{MSCcodes}

\section{Introduction: motivations, problem statement, and contributions}
\label{sec:intro}

This paper is a substantial extension of the 10-page conference version at NeurIPS 2022 \cite{widdowson2022resolving}.
The original paper introduced the Pointwise Distance Distribution (PDD) as an isometry invariant of a periodic set of points in any Euclidean space $\R^n$, and claimed the key properties (Lipschitz continuity, near-linear time computability, and generic completeness) in the fully periodic case,
where proofs in the appendices of \cite{widdowson2022resolving} were not expected to be reviewed.
This extended version defines PDD for any discrete set in a metric space and rigorously proves the properties above in finite and $l$-periodic cases for all $l\leq n$.
We adapt the invariants to a more convenient form, speed up the original implementation almost by two orders of magnitude, and report new experiments revealing duplicates in the world's largest materials databases.
\smallskip

\begin{figure}[h!]
\centering
\includegraphics[height=22.5mm]{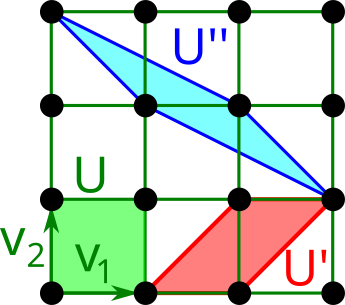}
\hspace*{0.5mm}
\includegraphics[height=22.5mm]{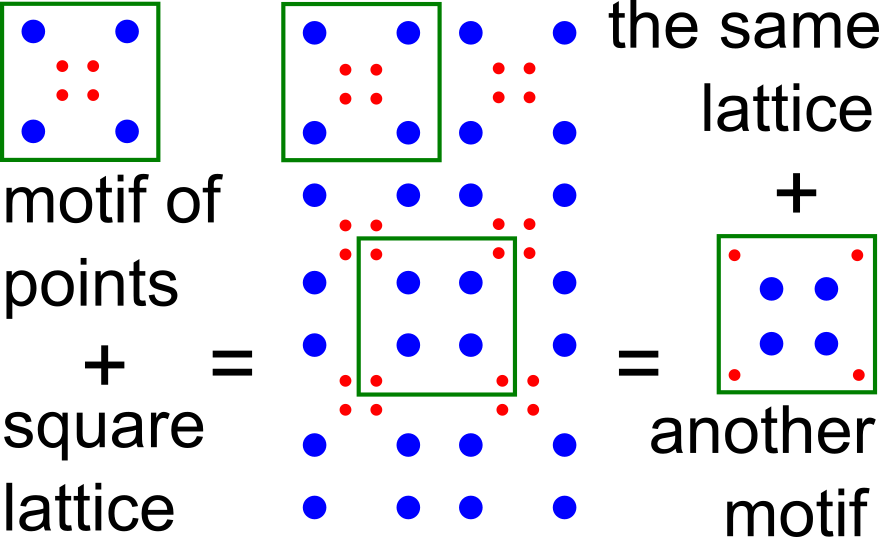}
\hspace*{0.5mm}
\includegraphics[height=22.5mm]{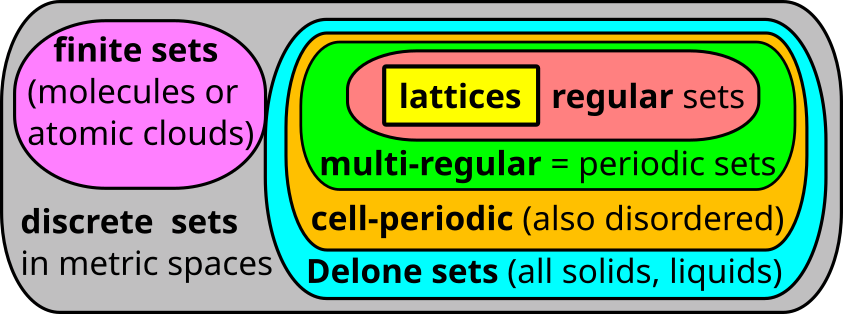}
\vspace*{-2mm}
\caption{\textbf{Left}: a lattice can be defined by many primitive bases.
\textbf{Middle}: a periodic set can be defined by different pairs (basis, motif).
\textbf{Right}: a hierarchy of discrete sets, which model periodic crystals and amorphous solids with points at atomic centers, see Definitions~\ref{dfn:metric},~\ref{dfn:periodic},~\ref{dfn:Delone_sets}, and~\ref{dfn:PPC}.  }
\label{fig:lattice_periodic_set_hierarchy}
\end{figure}

The continuous and generically complete invariants are motivated by the previously unresolved ambiguity of digital representations of molecules and crystals in terms of atomic coordinates or lattice bases.
Fig.~\ref{fig:lattice_periodic_set_hierarchy}~(middle) shows that the same periodic set can be obtained by periodically repeating different motifs of points. 

The crucial question ``same or different?'' was explicitly raised for crystals \cite{sacchi2020same} and makes sense for many other real objects.
For a cloud of unordered points in computer vision or chemistry applications, a list of atomic coordinates depends on a given coordinate system and an order of atoms.
The independence of coordinate representations is important for identifying rigid structures and rigid conformations of flexible molecules such as proteins whose properties depend on a rigid shape.  
\smallskip

Noisy measurements imply that any real objects are at least slightly different.
Hence the next practical question is ``how much different?''
If noise is ignored up to any positive threshold, noisy perturbations of atomic centers can be continued sufficiently long to make any given sets identical.  
This \emph{sorites paradox} \cite{hyde2011sorites} can be resolved by quantifying even tiny differences through a continuous distance metric.

\begin{dfn}[a \emph{discrete set} $S$ in a \emph{metric space} $X$ with a \emph{metric} $d_X$]
\label{dfn:metric}
A \emph{metric space} is any set $X$ of objects (called \emph{points}) with a \emph{distance metric} $d:X\times X\to\R$ satisfying the metric axioms: (1) \emph{coincidence} $d_X(a,b)=0$ if and only if $a=b$, (2) \emph{symmetry} $d_X(a,b)=d_X(b,a)$, and (3) \emph{triangle inequality} $d_X(a,b)+d_X(b,a)\geq d_X(a,c)$ for any points $a,b,c\in X$.
A set $S\subset X$ is called \emph{discrete} if there is a constant $\ep>0$ such that all points of $S$ are $\ep$-separated, so $d_X(a,b)\geq\ep$ for any $a,b\in S$. 
\end{dfn}

An example of a discrete set $S$ is a finite set in $\R^n$ with the Euclidean metric denoted by $|\vec p-\vec q|$ for any points $p,q\in\R^n$.
Here $\vec p$ denotes the vector from the origin $0\in\R^n$ to $p$.
The positivity $d_X(a,b)\geq 0$ follows from other axioms: $2d_X(a,b)=d_X(a,b)+d_X(b,a)\geq d_X(a,a)=0$.
Without the first axiom, $d$ is called a \emph{pseudo-metric} and can be the zero function: $d_X(a,b)=0$ for all $a,b$.
If the triangle inequality is allowed to fail with any additive error $\ep>0$, the results of clustering such as $k$-means and DBSCAN can be predetermined and hence may not be trustworthy \cite{rass2024metricizing}.

\begin{dfn}[lattice, unit cell, motif, $l$-periodic set] 
\label{dfn:periodic}
Vectors $\vec v_1,\dots,\vec v_n\in\R^n$ form a \emph{basis} if any vector in $\R^n$ can be written as $\vec v=\sum\limits_{i=1}^n x_i\vec  v_i$ for unique $x_1,\dots,x_n\in\R$. 
For $1\leq l\leq n$, the vectors $\vec v_1,\dots,\vec v_l$ define the \emph{lattice} $\La=\{\sum\limits_{i=1}^l c_i\vec  v_i \mid c_1,\dots,c_l\in\Z\}$ and the \emph{unit cell} $U=\{\sum\limits_{i=1}^n x_i\vec  v_i \mid x_1,\dots,x_l\in[0,1), x_{l+1},\dots,x_n\in\R\}\subset\R^n$.
If $l=n$, then $U$ is an $n$-dimensional parallelepiped.
If $l<n$, then $U$ is an infinite slab over an $l$-dimensional parallelepiped on $\vec v_1,\dots,\vec v_l$.   
For any finite set of points (called a \emph{motif}) $M\subset U$, the sum $S=M+\La=\{\vec p+\vec v \mid p\in M, v\in\La\}$ is an \emph{$l$-periodic point set}.
\end{dfn}

Any unit cell $U$ includes only a partial boundary: we exclude the points with any coefficient $t_i=1$, $i=1,\dots,l$, for convenience.
Then $\R^n$ for $l=n$ is tiled by the shifted cells $\{U+\vec v \mid \vec v\in\La\}$ without overlaps.
Any lattice is an example of a periodic set with one point in a motif.
Any periodic point set $S=M+\La$ can be considered a finite union $\bigcup_{p\in M}(\vec p+\La)$ of lattices whose origins are shifted to all $p\in M=S\cap U$.
\smallskip

If we double a unit cell in one direction, e.g. by taking the basis $2\vec v_1,\vec v_2,\dots,\vec v_n$, the doubled motif $M\cup(M+\vec v_1)$ with the sublattice on the new basis defines the original periodic point set $S=M+\La$.
A basis and its cell $U$ of $S$ are called \emph{primitive} if $S\cap U$ has the smallest size among all unit cells $U$ of $S$.
Fig.~\ref{fig:lattice_periodic_set_hierarchy}~(left) shows a square lattice in $\R^2$, which (as any lattice) can be generated by infinitely many primitive bases.
Even if we fix a basis, Fig.~\ref{fig:lattice_periodic_set_hierarchy}~(middle) shows that different motifs in the same primitive cell $U$ define equivalent periodic sets, which differ only by translation. 
\smallskip

Finite and periodic point sets represent molecules and periodic crystals at the atomic scale by considering zero-sized points at all atomic centers.
Chemical bonds can be modelled by straight-line edges between atomic centers.
However, even the strongest covalent bonds within a molecule depend on various thresholds for distances and angles.
So these bonds are not real sticks and only abstractly represent inter-atomic interactions, while atomic nuclei are real objects.
We model all materials at the fundamental level of atoms, which will suffice for all real materials.
Since any object can be defined in many different ways, 
Definition~\ref{dfn:equivalence} formalizes an equivalence.

\begin{dfn}[equivalence relation]
\label{dfn:equivalence}
An \emph{equivalence} is a binary relation (denoted by $\sim$) on any kind of objects satisfying the following axioms:
(1) \emph{reflexivity}: any objects $S$ is equivalent to itself, so $S\sim S$;
(2) \emph{symmetry}: if $S\sim Q$, then $Q\sim S$;
(3) \emph{transitivity}: if $S\sim Q$ and $Q\sim T$, then $S\sim T$.
Any object $S$ defines its \emph{equivalence class} $[S]=\{Q \mid Q\sim S\}$ as the full collection of all objects $Q$ equivalent to $S$.
\end{dfn}

The transitivity axiom justifies that all equivalence classes are disjoint: if $[S]$ and $[T]$ share a common object $Q$, then $[S]=[T]$.
Any well-defined classification should be based on an equivalence, whose practical examples are considered below.

\begin{dfn}[isometry, rigid motion in $\R^n$]
\label{dfn:isometry}
In a metric space $X$, an \emph{isometry} is any map $f:X\to X$ that preserves inter-point distances, i.e. $d(f(p),f(q))=d(p,q)$ for all $p,q\in X$.
In $\R^n$, any isometry decomposes into translations, rotations, and reflections, which generate the Euclidean group $\Eu(n)$.
If reflections are excluded, orientation-preserving isometries are also called \emph{rigid motions} and form 
group $\SE(n)$.  
\end{dfn}

Rigid motion (denoted by $\cong$) is the strongest equivalence for many objects in practice because translations and rotations of a molecule or solid material keep all their properties at least under the same ambient conditions such as temperature and pressure.
The isometry (denoted by $\simeq$) is only slightly weaker by allowing reflections.
Taking compositions with a uniform scaling in $\R^n$ or including (say) affine transformations gives weaker equivalences that define smaller spaces of classes.
\smallskip

This paper focuses on isometry as a more general equivalence defined in any metric space.
Our main problem will be to continuously parametrize equivalence classes of (various kinds of) discrete sets under isometry. 
Delone sets were introduced by B.~Delone \cite{delone1976local} as $(r,R)$-systems in $\R^n$ and make sense in any metric space $X$. 
Let $\bar B(p;r)=\{q\in X \mid d(p,q)\leq r\}$ be the closed ball with a center $p\in X$ and a radius $r$.

\begin{dfn}[Delone sets and $m$-regular sets]
\label{dfn:Delone_sets}
In a metric space $X$, a \emph{Delone} set $S$ is any subset of $X$ satisfying the following conditions:
\smallskip

\noindent
(a) \emph{packing}: 
there is a radius $r>0$ such that the closed balls $\bar B(p;r)$ for all points $p\in S$ are disjoint or, equivalently, all distances between points of $S$ are at least $2r$;
\smallskip

\noindent
(b) \emph{covering}: 
there is a radius $R>0$ such that $\bar B(p;R)$ for all $p\in S$ cover $X$, i.e. $\bigcup\limits_{p\in S} \bar B(p;R)=X$, or, equivalently, $\bar B(p;R)$ for any $p\in X$ has at least one point of $S$.
\smallskip

\noindent
A Delone set is called \emph{$m$-regular} if $S$ splits into $m$ classes under the \emph{global isometry equivalence}: $p\sim q$ if there is an isometry $f:X\to X$ such that $f(S)=S$, $f(p)=q$.
\end{dfn}

The packing condition implies that $S$ is a discrete set in $X$ by specifying a minimum inter-point distance $\ep=2r$ and is well-motivated by the fact that real atoms strongly repel each other at very short distances \cite{feynman2011lectures}.
The covering condition says that $X$ has no unbounded `empty' balls without any points of $S$ and is also motivated by the absence of infinite round pores in solid materials, liquids, and dense gases.
\smallskip

All $m$-regular sets for $m>1$ are also called \emph{multi-regular}, while 1-regular sets are often called \emph{regular}.  
Any lattice $\La\subset\R^n$ is regular because the required isometry $f:\La\to\La$ mapping a point $p\in\La$ to another $q\in\La$ is the translation by the vector $\vec q-\vec p$.
Similarly, any periodic point set $S$ is $m$-regular, where $m$ is upper bounded by the size of a motif $M$ of $S$. 
A honeycomb periodic set in $\R^2$ modeling graphene is regular, but not a lattice because there are two points in a primitive unit cell.
The regularity means that $S$ looks the same when viewed from any point of $S$.
Fig.~\ref{fig:lattice_periodic_set_hierarchy}~(middle) shows a 2-regular set whose points split into red and blue classes under the global isometry equivalence.
\cite[Theorem~1.3]{dolbilin1998multiregular} proved that any multi-regular Delone set is periodic.
\smallskip

A finite set in $\R^n$ is not a Delone set but any finite subset of a finite metric space is Delone.
The latter special case is indicated by cyan and magenta regions slightly touching each other in Fig.~\ref{fig:lattice_periodic_set_hierarchy}~(middle).
All other inclusions are strict, not to scale.
\smallskip

The key tool in classifying under an equivalence 
is an \emph{invariant} that is a function $I$ taking the same value on all equivalent objects.
For a finite set $S\subset\R^n$, the number $m$ of points is an isometry invariant, but the geometric average $\dfrac{1}{m}\sum\limits_{p\in S}\vec p$ is not. 
\smallskip

We state the mapping problem for any discrete sets under isometry, though the same conditions make sense for many other objects, e.g. graphs and polygonal meshes, and equivalences, e.g. rigid motions, affine or projective transformations in $\R^n$.

\begin{pro}[\textbf{geo-mapping problem} for any discrete sets under isometry]
\label{pro:map}
For a metric space $X$ with a metric $d_X$, find a map $I:\{$discrete sets of unordered points in $X\}\to$ a metric space with a metric $d$ satisfying the following conditions.
\smallskip

\noindent{(a)} 
\textbf{Completeness}: 
any sets $S\simeq Q$ are isometric in $X$ if and only if $I(S)=I(Q)$.
\smallskip

\noindent{(b)} 
\textbf{Realizability}: 
the image $\{I(S) \mid S\subset X\}$ is parametrized so that taking any value of $I$ from this image allows us to reconstruct $S\subset X$ uniquely under isometry of $X$. 
\smallskip

\noindent{(c)} 
\textbf{Lipschitz continuity}: 
there is a constant $\la$ such that if $Q$ is obtained by perturbing each point of $S$ up to any $\ep$ in the metric $d_X$, then $d(I(S),I(Q))\leq \la\ep$.  
\smallskip

\noindent{(d)} 
\textbf{Computability}: 
the invariant $I$, the metric $d$, and the reconstruction of $S\subset X$ from $I(S)$ can be computed in a time that depends polynomially on the input sizes. 
\end{pro} 
 
For any finite set $S\subset X$, its input size is the number $m$ of points.
For any periodic point set $S\subset\R^n$, its input size is the number $m$ of points in a motif $M$ from Definition~\ref{dfn:periodic} because a Crystallographic Information File (CIF) specifying a basis and atomic coordinates in this basis has a linear length $O(m)$ in the motif size $m$.
Some infinite Delone sets can described in a finite form, e.g. certain aperiodic crystals \cite{senechal1996quasicrystals} can be obtained as projections of periodic crystals in higher dimensions.
\smallskip

We leave these general cases for future work and focus on finite and periodic point sets, which already cover many applications where Problem~\ref{pro:map} was widely open.

\begin{figure}[h!]
\includegraphics[height=19mm]{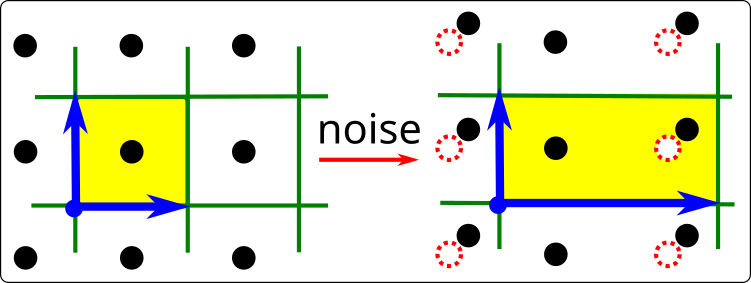}
\hspace*{0.5mm}
\includegraphics[height=19mm]{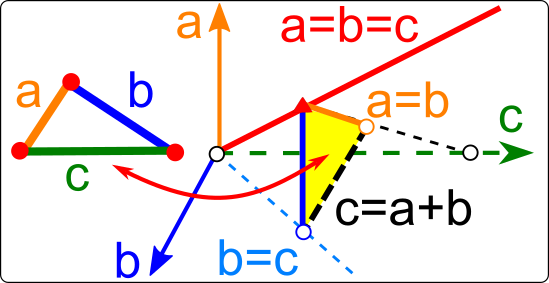}
\hspace*{0.5mm}
\includegraphics[height=19mm]{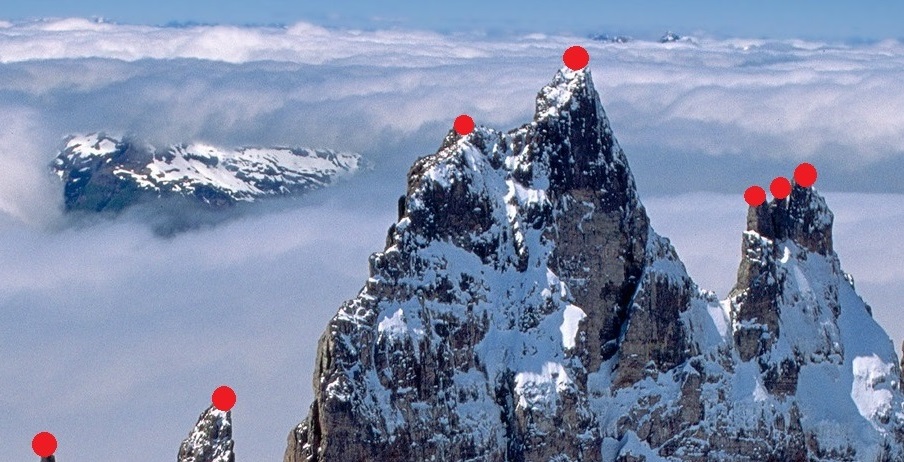}
\vspace*{-4mm}
\caption{
\textbf{Left}: the symmetry group and a reduced cell discontinuously change under tiny noise.
\textbf{Middle}:  the space of 3 points under isometry is parametrized by inter-point distances $0<a\leq b\leq c\leq a+b$. 
\textbf{Right}: energy landscapes of crystals show optimized structures as isolated peaks of height$=-$energy.
To see beyond the `fog', we need a map parametrized by invariants in Problem~\ref{pro:map}.}
\label{fig:noise_double_triangles_peaks}
\end{figure}

The completeness in (\ref{pro:map}a) implies that the invariant $I$ is a descriptor with \emph{no false negatives} and \emph{no false positives} for all discrete sets, and hence can be considered a DNA-style code that uniquely identifies any isometry class.
The realizability in (\ref{pro:map}b) is even stronger and enables us to sample the space of realizable invariants and reconstruct the resulting set $S$, while a real DNA code is insufficient to grow a living organism.
The Lipschitz continuity in (\ref{pro:map}c) is motivated by ever-present thermal vibrations and experimental noise.
Fig.~\ref{fig:noise_double_triangles_peaks}~(left) shows that almost any perturbation of points can arbitrarily scale up a primitive cell.
This inherent discontinuity of traditional cell-based representations remained a practical loophole in crystallography at least since 1965 \cite{lawton1965reduced} and allowed disguising known materials by a slight perturbation changing the space group and even the primitive cell volume, and also by replacing some chemical elements to avoid detection by chemical composition \cite[section~6]{anosova2024importance}.
\smallskip

Fig.~\ref{fig:noise_double_triangles_peaks}~(middle) shows a solution of Problem~\ref{pro:map} for $m=3$ points 
saying that any triangle is determined under isometry by 3 ordered inter-point distances. 
Real or simulated crystals are local optima (mountain peaks) in Fig.~\ref{fig:noise_double_triangles_peaks}~(right) on a continuous space of (isometry classes of) periodic point sets, whose `geography' was unknown. 
\smallskip

\noindent
\textbf{Contributions}.
We introduce the Pointwise Distance Distribution for any discrete set in a metric space. 
This generality is of broad interest to experts in computational geometry and applications to physical objects from molecules to solid or even liquid materials.
The previously unpublished aspects are the asymptotic for $l$-periodic sets, rigorous proofs of the Lipschitz continuity (also for adjusted and normalized invariants), near-linear time computability, and generic completeness in the finite and periodic case.
The linear-time algorithms and the hierarchical nature of PDD computations have become extremely important for big databases, especially in the last years when millions of artificial structures were claimed `new' without checking for duplication with known crystals.  
The decisive advance is closing this discontinuity loophole in crystallography, which is demonstrated for the world's largest databases.

\section{Review of rigorous approaches to mapping spaces of discrete sets}
\label{sec:review}

This section reviews progress in solving Problem~\ref{pro:map} for finite and periodic point sets by proof-based methods, not by experimental studies, which are reviewed in \cite{widdowson2022resolving, widdowson2023recognizing}. 
Finite sets have two subcases: ordered points (easy) and unordered (much harder). 
\smallskip

\noindent
\textbf{Ordered finite sets}.
Kendall's shape theory \cite{kendall2009shape} studies ordered points $p_1,\dots,p_m\in\R^n$ whose complete isometry invariant is the matrix of distances \cite{schoenberg1935remarks} 
or scalar products $\vec p_i\cdot\vec  p_j$ \cite[chapter 2.9]{weyl1946classical}, see a linear-time invariant in \cite{anosova2025complete,wlodawer2025duplicate}.
A brute-force extension to $m$ unordered points requires $m!$ permutations of points, ruled out by (\ref{pro:map}d).
\smallskip

\begin{figure}[h!]
\centering
\includegraphics[height=20.5mm]{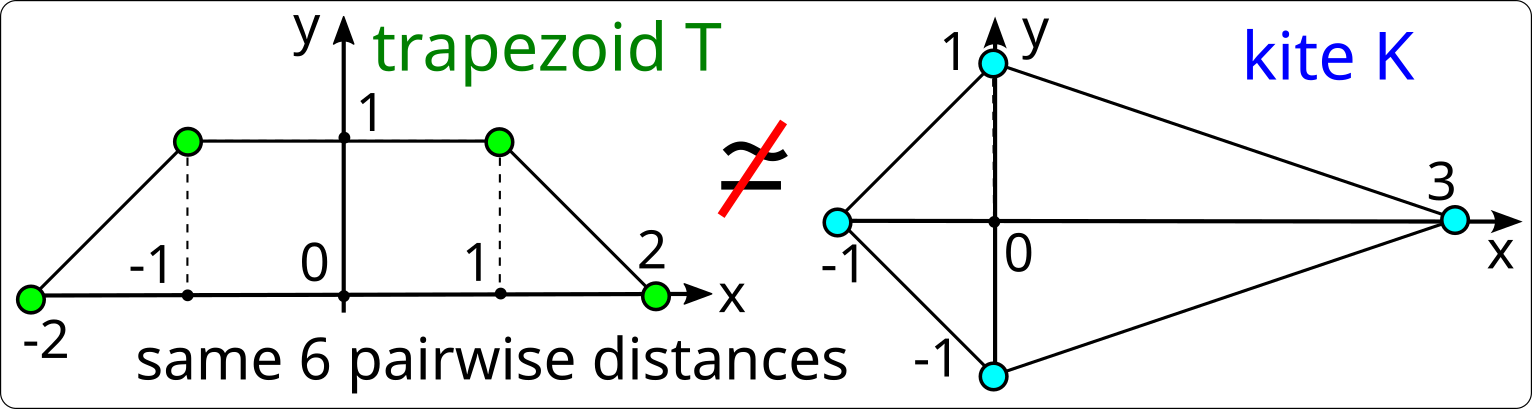}
\hspace*{0.5mm}
\includegraphics[height=20.5mm]{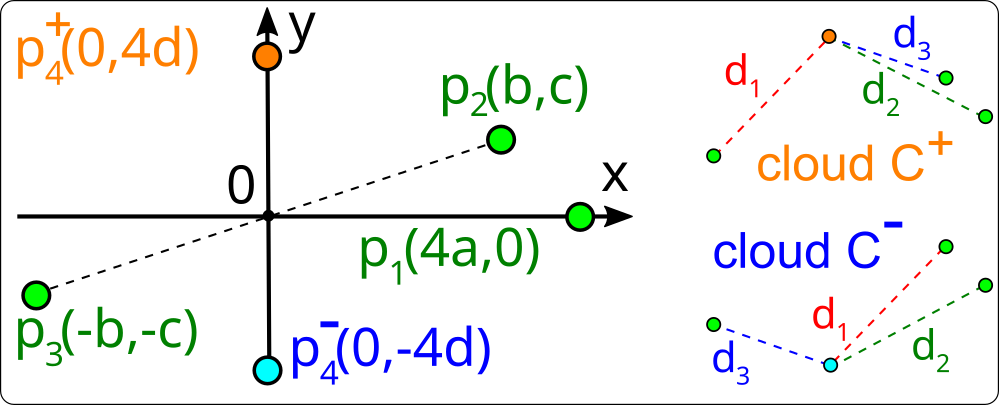}
\caption{
Non-isometric clouds of 4 points with the same 6 pairwise distances.
\textbf{Left}: the trapezoid $T$ has points $(\pm 1,1)$, $(\pm 2,0)$.
The kite $K$ has points $(3,0)$, $(-1,0)$, $(0,\pm 1)$.
\textbf{Right}:  the infinite family of non-isometric clouds $C^+\not\simeq C^-$ sharing $p_1,p_2,p_3$ and depending on parameters $a,b,c,d>0$ \cite{caelli1979generating}.}
\label{fig:4-point_clouds}
\end{figure}

\noindent
\textbf{Unordered finite sets} (point clouds).
Extending the case of $m=3$ points in Fig.~\ref{fig:noise_double_triangles_peaks}~(middle), Boutin and Kemper proved in 2004  that the unordered distribution of distances between $m$ points uniquely determines a generic $m$-point cloud $C\subset\R^n$ under isometry \cite{boutin2004reconstructing}.
The genericity condition allows almost all clouds apart from a measure 0 subspace among all clouds.
For any cloud $C$ of $m$ unordered points in a metric space $X$, the vector $\SPD(C)$ consists of $\frac{m(m-1)}{2}$ \emph{Sorted Pairwise Distances} written in increasing order and computable in time $O(m^2\log m)$.
The space of 4-point clouds in $\R^2$ has dimension 5 because 6 inter-point distances satisfy one polynomial equation saying that the tetrahedron on these points has volume 0.
Fig.~\ref{fig:4-point_clouds} shows a 4-parameter family of pairs of non-isometric clouds with the same vector $\SPD$.
\smallskip

Problem~\ref{pro:map} expands the question `Can we hear the shape of a drum?' \cite{kac1966can} which has the negative answer in terms of 2D polygons that are indistinguishable by spectral invariants \cite{gordon1992isospectral,gordon1992one,reuter2006laplace,cosmo2019isospectralization,marin2021spectral}.
Problem~\ref{pro:map} looks for stronger invariants that can completely `sense' as in (\ref{pro:map}b), not only `hear', the rigid shape of any cloud.
\smallskip
 
\noindent
\textbf{Computational geometry} studied weaker versions of Problem~\ref{pro:map} by developing canonical representations of point clouds \cite{alt1988congruence,brass2004testing}, which can be considered complete invariants, and also metrics between isometry classes of clouds. 
For example, any metric between fixed clouds extends to their isometry classes \cite{chew1999geometric} by minimization over infinitely many transformations from the group $\Eu(n)$.
This extension of the Hausdorff distance \cite{hausdorff1919dimension} for $m$-point clouds in $\R^2$ has time $O(m^5\log m)$, see \cite{goodrich1999approximate}. 
The Gromov-Wasserstein metrics \cite{memoli2011gromov,memoli2021gromov} are defined for any metric-measure spaces also by minimizing over infinitely many correspondences between points, but cannot be approximated with a factor less than 3 in polynomial time unless P=NP, see Corollary~3.8 in \cite{schmiedl2017computational} and polynomial algorithms for partial cases in \cite{agarwal2018computing,majhi2024approximating}.
\smallskip

\noindent
Computing a metric between isometry classes of clouds is only a small part of Problem~\ref{pro:map}.
Indeed, to efficiently navigate on Earth, in addition to distances between cities, we need a satellite-type view of the full planet and hence a realizable continuous invariant $I$, which can be used as the latitude and longitude coordinates.
\smallskip

\noindent
\textbf{Geometric Data Science}\cite{anosova2025geometric,anosova2026seeing} gradually solved simpler versions of Problem~\ref{pro:map} since 2020 when the continuity was first stated for lattices \cite{mosca2020voronoi} and then for general periodic sets \cite{anosova2021introduction,edelsbrunner2021density}.
The case of 2D lattices was finished in \cite{kurlin2024mathematics} with a weaker H\"older continuity for a stronger relation under rigid motion, because the Lipschitz continuity is impossible for perturbations of a lattice basis.
See continuous chiral distances and geographic-style maps in \cite{bright2023continuous,bright2023geographic}, and
 complete invariants of 3D lattices in \cite{kurlin2022complete,bright2021complete}.
\smallskip

The Pointwise Distance Distribution (PDD) solved Problem~\ref{pro:map} for finite unordered sets with distinct distances in $\R^n$ \cite[Theorem~16]{widdowson2022resolving}.
This PDD appeared as a local distribution of distances in the finite case \cite{memoli2011gromov} and was extended to higher order invariants in the periodic case \cite{widdowson2025higher}.
For all finite sets in $\R^n$, \cite{widdowson2023recognizing,kurlin2024polynomial} developed complete invariants under rigid motion with Lipschitz continuous metrics.
For periodic sets, the complete \emph{isoset} \cite{anosova2021isometry} has only an approximate algorithm for a continuous metric \cite{anosova2026recognition,mcmanus2025computing}.
For generic 1-periodic sets, \cite{kurlin2025complete} defined an exact and continuous metric. 
 
\section{The Pointwise Distance Distribution and other isometry invariants}
\label{sec:invariants}

This section introduces the Pointwise Distance Distribution (PDD) for any discrete set $S$ with a specified finite subset $M$ in a metric space $X$.
If $S$ is finite, we set $M=S$.
If $S$ is periodic, $M$ is a motif of $S$, but $\PDD$ will depend only on $S$, not on $M$.

\begin{dfn}[PDD and AMD invariants]
\label{dfn:PDD}
Let $M=\{p_1,\dots,p_m\}$ be a finite subset of a discrete set $S$ in a metric space $X$.
Fix an integer $k\geq 1$.
For every point $p\in M$, let $d_1(p)\leq\dots\leq d_k(p)$ be the distances from $p$ to its $k$ nearest neighbors within the full set $S$ (not restricted to $M$).
The matrix $D(S,M;k)$ has $m$ rows consisting of the distances $d_1(p_i),\dots,d_k(p_i)$ for $i=1,\dots,m$.
If any $l\geq 2$ rows coincide, we collapse them into a single row with the weight $l/m$.
The resulting unordered set (written as a matrix) of maximum $m$ rows and 
$k+1$ columns, including the extra column of weights, is the \emph{Pointwise Distance Distribution} $\PDD(S,M;k)$. 
The \emph{Average Minimum Distance} $\AMD_i$ is the weighted average of the $i$-th column in $\PDD(S,M;k)$ for each $i=1,\dots,k$.
Let $\AMD(S,M;k)$ denote the vector $(\AMD_1,\dots,\AMD_k)$. 
\end{dfn}

Definition~\ref{dfn:PDD} introduced the isometry invariant 
$\PDD(S,M;k)$ of a pair $(S,M)$ for a finite subset $M$ in any Delone set $S$.
For any $l$-periodic point set $S\subset\R^n$, Theorem~\ref{thm:invariance} will prove that $\PDD$ is independent of a motif $M\subset S$.
We use the simpler notations $\PDD(S;k), \AMD(S;k)$ in the finite ($S=M$) and periodic cases.

\begin{exa}[4-point clouds $T,K$ in {Fig.~\ref{fig:4-point_clouds}~(left)}]
\label{exa:4-points}
Table~\ref{tab:ordered_distances_T+K} shows the $4\times 3$ matrices $D(S;3)$ from Definition~\ref{dfn:PDD}.
The matrix $D(T;3)$ in Table~\ref{tab:ordered_distances_T+K} has two pairs of identical rows, so the matrix $\PDD(T;3)$ consists of two rows of weight $\frac{1}{2}$ below.
The matrix $D(K;3)$ in Table~\ref{tab:ordered_distances_T+K} has only one pair of identical rows, so $\PDD(K;3)$ has three rows of weights $\frac{1}{2}$, $\frac{1}{4}$, $\frac{1}{4}$.
Then $T,K$ are distinguished by $\PDD$s even for $k=1$.

\begin{table}[h!]
\caption{Each point of $T,K\subset\R^2$ in Figure~\ref{fig:4-point_clouds}~(left) has distances to other points in increasing order.
After keeping only distances (not neighbors), the resulting $\PDD$s distinguish $T\not\simeq K$, see Example~\ref{exa:4-points}.}
\label{tab:ordered_distances_T+K}
\centering
\begin{tabular}{lccc}
points of $T$ & dist. to neighbor 1 & dist. to neighbor 2 & dist. to neighbor 3 \\
\hline
$(-2,0)$ & $\sqrt{2}$ to $(-1,+1)$ & $\sqrt{10}$ to $(+1,+1)$  & $4$ to $(+2,0)$ \\
$(+2,0)$ & $\sqrt{2}$ to $(+1,+1)$ & $\sqrt{10}$ to $(-1,+1)$  & $4$ to $(-2,0)$ \\
$(-1,+1)$ & $\sqrt{2}$ to $(-2,0)$ & $2$ to $(+1,+1)$ & $\sqrt{10}$ to $(+2,0)$ \\
$(+1,+1)$ & $\sqrt{2}$ to $(+2,0)$ & $2$ to $(-1,+1)$ & $\sqrt{10}$ to $(-2,0)$  
\end{tabular}
\smallskip

\begin{tabular}{lccc}
points of $K$ & dist. to neighbor 1 & dist. to neighbor 2 & dist. to neighbor 3  \\
\hline
$(-1,0)$ & $\sqrt{2}$ to $(0,-1)$ & $\sqrt{2}$  to $(0,+1)$ & $4$ to $(3,0)$ \\
$(+3,0)$ & $\sqrt{10}$ to $(0,-1)$ & $\sqrt{10}$ to $(0,+1)$ & $4$ to $(-1,0)$ \\
$(0,-1)$ & $\sqrt{2}$ to $(-1,0)$ & $2$ to $(0,+1)$ & $\sqrt{10}$ to $(3,0)$\\
$(0,+1)$ & $\sqrt{2}$ to $(-1,0)$ & $2$ to $(0,-1)$ & $\sqrt{10}$ to $(3,0)$
\end{tabular}
\end{table}

\vspace*{-4mm}
\noindent
$\PDD(T)=\left(\begin{array}{c|ccc}
1/2 & \sqrt{2} & 2 & \sqrt{10} \\
1/2 & \sqrt{2} & \sqrt{10}  & 4
\end{array}\right)\neq 
\PDD(K)=\left(\begin{array}{c|ccc}
1/4 & \sqrt{2} & \sqrt{2}  & 4 \\
1/2 & \sqrt{2} & 2 & \sqrt{10} \\
1/4 & \sqrt{10} & \sqrt{10}  & 4
\end{array}\right)$.
\end{exa}

Theorem~\ref{thm:invariance} extends \cite[Theorem~3.2, proved in appendix~C]{widdowson2022resolving} from $n$-periodic sets to all finite sets, $l$-periodic sets, and arbitrary pairs $(S,M)$ from Definition~\ref{dfn:PDD}.

\begin{thm}[invariance]
\label{thm:invariance}
\textbf{(a)} 
For any isometry $S\to Q$ of discrete sets that maps a finite subset $M\subset S$ of $m$ points to $N\subset Q$, we have $\PDD(S,M;k)=\PDD(Q,N;k)$ and $\AMD(S,M;k)=\AMD(Q,N;k)$ for $1\leq k<m$. 
Hence, if $S=M$ is discrete, then $\PDD(S;k)$ and $\AMD(S;k)$ are 
isometry invariants of $S$.
\smallskip

\noindent
\textbf{(b)} 
For any $l$-periodic point set $S\subset\R^n$, where $1\leq l\leq n$, 
$\PDD(S;k)$ and $\AMD(S;k)$ are invariants of $S$ (independent of a motif $M\subset S$) under isometry of $\R^n$ for $k\geq 1$.
\end{thm}
\begin{proof}
\textbf{(a)} 
For any sets $M\subset S$ and their isometric images $N\subset Q$, the invariance follows from the fact that any isometry preserves all inter-point distances. 
\smallskip

\noindent
\textbf{(b)} 
For any $l$-periodic point set $S=\La+M\subset\R^n$, we first show that scaling up a cell $U$ and hence the motif $M=S\cap U$ of $m$ points keeps $\PDD$ invariant.
For any integer $b\geq 1$, a matrix $B\in\GL(l;\Z)$ with $|\det B|=b$ acts on the first $l$ vectors $\vec v_1,\dots,\vec v_l$ that generate the $l$-dimensional base parallelepiped $P$ of $U$ in Definition~\ref{dfn:periodic}. 

Let $B(U)\subset\R^n$ denote the cell obtained from $U$ by applying $B$ to $P$ and keeping all other basis vectors $v_{l+1},\dots,v_n$ fixed.
Then $D(S,S\cap B(U);k)$ from Definition~\ref{dfn:PDD} has the larger size $bm\times k$ but (due to periodicity of $S$) splits into $m$ blocks, each corresponding to $b$ points of the scaled motif $S\cap B(U)$ that are obtained from a single point $p\in M$ by translations by vectors of $\La$.
Since translations preserve distances, each of $m$ blocks has $b$ identical rows of distances to $k$ neighbors in $S$, the same as in $D(S,M;k)$.
Then $\PDD(S,S\cap B(U);k)=\PDD(S,M;k)$ due to collapsing of identical rows
in Definition~\ref{dfn:PDD}.
So $\PDD(S;k)$ is independent of any motif $M=S\cap U$.

Now we prove that $\PDD(S;k)$ is preserved by any isometry $f$ of $\R^n$.
Any primitive cell $U$ of $S$ is bijectively mapped by $f$ to the unit cell $f(U)$ of $Q=f(S)$, which should be also primitive.
Indeed, if $Q$ is preserved by a translation along a vector $v$ that doesn't have all integer coefficients in the basis of $f(U)$, then $S=f^{-1}(Q)$ is preserved by the translation along $f^{-1}(v)$, which doesn't have all integer coefficients in the basis of $U$, so $U$ was non-primitive.
Since $U$ and $f(U)$ have the same number of points from $S$ and $Q=f(S)$, the isometry $f$ gives a bijection between the motifs of $S,Q$.

For any periodic sets $S,Q$, because $f$ maintains distances, every list of ordered distances from $p_i\in S\cap U$ to its first $k$ nearest neighbors in $S$ coincides with the list of the ordered distances from $f(p_i)$ to its first $k$ neighbors in $Q$.
These coincidences of distance lists give $\PDD(S;k)=\PDD(Q;k)$ after collapsing identical rows.  
\end{proof}


If we increase $k$, more columns with larger values are added to $\PDD(S;k)$ but all previous distances remain the same.
Definition~\ref{dfn:PPC} will help describe the asymptotic of $\PDD(S;k)$ as $k\to+\infty$ in Theorem~\ref{thm:asymptotic}, which uses Lemma~\ref{lem:cylinder_size} extending \cite[Lemma 11]{widdowson2022average} to $l$-periodic sets $S\subset\R^n$ for any $1\leq l\leq n$, see all skipped proofs in appendix~\ref{sec:PDD_ext_proofs}

\begin{dfn}[Point Packing Coefficient $\PPC$ of a cell-periodic set $S$]
\label{dfn:PPC}
For $1\leq l\leq n$ and a basis $\vec v_1,\dots,\vec v_n\in\R^n$, consider the
\emph{lattice} $\La=\{\sum\limits_{i=1}^l c_i\vec  v_i \mid c_1,\dots,c_l\in\Z\}$ and the unit cell $U=\{\sum\limits_{i=1}^n x_i\vec  v_i \mid x_1,\dots,x_l\in[0,1), x_{l+1},\dots,x_n\in\R\}$.
A discrete set $S\subset\R^n$ is \emph{cell-periodic} if $S$ has a fixed number $m$ points in every shifted cell $U+\vec v$ for all $\vec v\in\La$. 
If $l<n$, let $R^l\subset\R^n$ be the subspace spanned by $\vec v_1,\dots,\vec v_l$, then $U$ is an infinite slab based on the $l$-dimensional parallelepiped of volume $\vol[U\cap R^l]$ .
The volume of the unit ball in $\R^l$ is $V_l=\dfrac{\pi^{l/2}}{\Ga(\frac{l}{2}+1)}$, where Euler's Gamma function 
is 
$\Ga(m)=(m-1)!$ and $\Ga(\frac{m}{2}+1)=\sqrt{\pi}(m-\frac{1}{2})(m-\frac{3}{2})\cdots\frac{1}{2}$ for any integer $m\geq 1$.
Define the \emph{Point Packing Coefficient} of $S$ as $\PPC(S)=\sqrt[l]{\dfrac{\vol[U\cap R^l]}{mV_l}}$.
\end{dfn}

Any $l$-periodic set is cell-periodic, but all cell-periodic sets form a wider collection of Delone sets and model disordered solid materials that can have an underlying lattice with atoms at different positions in periodically translated cells $U+\vec v$, see Fig.~\ref{fig:lattice_periodic_set_hierarchy}.

\begin{lem}[bounds on points within a cylinder]
\label{lem:cylinder_size}
For any $1\leq l\leq n$ and a basis $\vec v_1,\dots,\vec v_n\in\R^n$, let $S\subset\R^n$ be a cell-periodic set with a unit cell $U$ based on the $l$-dimensional parallelepiped $U\cap R^l$, where $R^l\subset\R^n$ is spanned by $\vec v_1,\dots,\vec v_l$.
Define the \emph{width} $w$ of $U$ as $\sup\limits_{u,v\in U\cap R^l}|\vec u-\vec v|$.
For any point $p\in S\cap U$ and a radius $r$, consider the cylinder $C(p;r)=\{\sum\limits_{i=1}^n x_i\vec  v_i \text{ such that } x_{1},\dots,x_n\in\R
 \text{ and } |p-\sum\limits_{i=1}^l x_i\vec  v_i|\leq r \}\subset\R^n$,
\begin{align*}
& \text{the \emph{lower} union }
U^-(p;r)=\bigcup \{(U+\vec v) \text{ such that } \vec v\in\La, (U+\vec v)\subset C(p;r)\}\subset\R^n, \\
& \text{the \emph{upper} union }
U^+(p;r)=\bigcup \{(U+\vec v) \text{ such that }  \vec v\in\La, (U+\vec v)\cap  C(p;r)\neq\emptyset\}.
\end{align*}
Let the unions $U^\pm(p;r)$ contain $m^{\pm}(p;r)$ shifted cells of $U+\vec v$ for some $\vec v\in\La$. 
Let $S$ have $m=|S\cap U|$ points in $U$.
Then the number of points from $S$ in $C(p;r)$ satisfies
$$\left(\dfrac{r-w}{\PPC(S)}\right)^l\leq
m^{-}(p;r) m\leq |S\cap C(p;r)|\leq 
m^{+}(p;r)m\leq
\left(\dfrac{r+w}{\PPC(S)}\right)^l.$$
\end{lem}


\begin{lem}[distance bounds]
\label{lem:distance_bounds}
In the notations of Lemma~\ref{lem:cylinder_size}, let the subspace $R^{n-l}$ be orthogonal to $R^l$, which is spanned by the first $l$ basis vectors of a cell $U$.
Let the \emph{height} $h$ of a cell-periodic set $S\subset\R^n$ with the cell $U$ be the maximum distance between points in the orthogonal projection of $S$ to $R^{n-l}$, so if $l=n$, then $h=0$.
For any point $p\in S\cap U$, let $d_k(S;p)$ be the distance from $p$ to its $k$-th nearest neighbor in the full set $S$.
Then $\PPC(S)\sqrt[l]{k}-w < d_k(S;p) \leq 
\sqrt{(\PPC(S)\sqrt[l]{k}+w)^2+h^2}$, $k\geq 1$.
\end{lem}

\begin{thm}[asymptotic of $\PDD(S;k)$ as $k\to+\infty$]
\label{thm:asymptotic}
For any point $p$ in a cell-periodic set $S\subset\R^n$, let $d_k(S;p)$ be the distance from $p$ to its $k$-th nearest neighbor in $S$.
Then  
$\lim\limits_{k\to+\infty}\dfrac{d_k(S;p)}{\sqrt[l]{k}}=\PPC(S)$ and hence
$\lim\limits_{k\to+\infty}\dfrac{\AMD_k(S)}{\sqrt[l]{k}}=\PPC(S)$.
\end{thm}
\begin{proof}[Proof of Theorem~\ref{thm:asymptotic}]
Lemma~\ref{lem:distance_bounds} gives the following bounds for 
$\de_k=\dfrac{d_k(S;p)}{\sqrt[l]{k}}-\PPC(S)$.
The lower bound is $\de_k>-u_k$, where $u_k=\dfrac{w}{\sqrt[l]{k}}\to 0$ as $k\to+\infty$ because $w$ is fixed. 
The upper bound is $\de_k\leq \sqrt{(\PPC(S)+u_k)^2+(h/\sqrt[l]{k})^2}-\PPC(S)\to 0$ as $k\to+\infty$, because $h$ is fixed.
Hence $\de_k=\dfrac{d_k(S;p)}{\sqrt[l]{k}}-\PPC(S)\to 0$ as $k\to+\infty$.
\end{proof}

\smallskip

By Theorem~\ref{thm:asymptotic}, $\AMD_k(S)$ and all distances in the last column of $\PDD(S;k)$ asymptotically approach $\PPC(S)\sqrt[l]{k}$ as $k\to+\infty$ and hence are mainly determined by $\PPC(S)$ for large $k$.
That is why the most descriptive information is contained in $\PDD(S;k)$ for smaller values of $k$, e.g. we use $k=100$ atomic neighbors in most experiments on crystals.
To neutralize the asymptotic growth, we subtract and also normalize by the term $\PPC(S)\sqrt[l]{k}$ to get simpler invariants under uniform scaling.

\begin{dfn}[simplified invariants $\ADA$, $\PDA,\AND$, $\PND$]
\label{dfn:ADA}
Let $S\subset\R^n$ be any $l$-periodic set with an underlying lattice generated by $l$ vectors. 
The \emph{Average Deviation from Asymptotic} is $\ADA_k(S)=\AMD_k(S)-\PPC(S)\sqrt[l]{k}$ for $k\geq 1$.
The \emph{Pointwise Deviation from Asymptotic} $\PDA(S;k)$ is obtained from the matrix $\PDD(S;k)$ by subtracting $\PPC(S)\sqrt[l]{j}$ from any distance in a row $i$ and a column $j$ for $i\geq 1\leq j\leq k$.
The \emph{Average Normalized Deviation} is $\AND_k(S)=\ADA_k(S)/(\PPC(S)\sqrt[l]{k})$, $k\geq 1$.
The \emph{Pointwise Normalized Deviation} $\PND(S;k)$ obtained from $\PDA(S;k)$ by dividing every element in a row $i$ and a column $j$ by $\PPC(S)\sqrt[l]{j}$ for $i\geq 1\leq j\leq k$.
\end{dfn} 

\begin{cor}[invariance of $\AND,\PND$ under uniform scaling]
\label{cor:AND}
For any $l$-periodic set $S\subset\R^n$, $\AND_k(S)$ and $\PND(S;k)$ in Definition~\ref{dfn:ADA} are invariant under isometry and uniform scaling for any $k\geq 1$.
Moreover, $\AND_k(S)\to 0$ as $k\to+\infty$.
\end{cor}
\begin{proof}
By Theorem~\ref{thm:invariance}, $\PDD(S;k)$ and hence all deviations in Definition~\ref{dfn:ADA} are invariant under isometry.
Under uniform scaling $p\mapsto cp$ for a real constant $c\neq 0$, any inter-point distance and $\PPC(S)=\sqrt[l]{\dfrac{\vol[U\cap R^l]}{mV_l}}$ is multiplied by $c$ because $\vol[U\cap R^l]$ is scaled by the factor $c^l$.
Hence $\AND_k(S)$ and $\PND(S;k)$ are invariant under both isometry and uniform scaling.  
To prove that $\AND_k(S)\to 0$ as $k\to+\infty$, use Theorem~\ref{thm:asymptotic}: 
$\AND_k(S)=\dfrac{\ADA_k(S)}{\PPC(S)\sqrt[l]{k}}=\dfrac{\AMD_k(S)}{\PPC(S)\sqrt[l]{k}}-1\to \dfrac{\PPC(S)}{\PPC(S)}-1=0$.
\end{proof}

We conjecture that $\ADA_k(S)\to 0$ as $k\to+\infty$ without the extra division by $\sqrt[l]{k}$ for $l\geq 2$, which is confirmed by experiments on crystals 
and holds for $S=\Z^n$ by Example~\ref{exa:cubic_lattice_asymptotic}. 
\smallskip

The key input sizes for computing $\PDD(S;k)$ of any $l$-periodic point set $S\subset\R^n$ are the number $m$ of points in a unit cell $U$ and the number $k$ of neighbors.
The full input consists of $k$, a basis of $U$ and a motif of $m$ points with coordinates in this basis as described in Definition~\ref{dfn:periodic}. 
For a fixed dimension $n$ and other parameters, the asymptotic complexity of $\PDD(S;k)$ will depend near linearly on both $k,m$.
\smallskip
 
The output $\PDD(S;k)$ is a matrix with at most $m$ rows and exactly $k+1$ columns, where $m$ is the number of motif points. 
The first column contains the weights of rows, which sum to 1 and are proportional to the number of appearances of each row before collapsing in Definition~\ref{dfn:PDD}, see Python code in appendix~\ref{sec:PDD_ext_code}.
A different early version of Theorem~\ref{thm:algorithm} had a 5-line proof in \cite[Theorem 5.1, appendix~C]{widdowson2022resolving}.

\begin{thm}[$\PDD$ complexity]
\label{thm:algorithm}
Let $S\subset\R^n$ be any $l$-periodic set with a minimum inter-point distance $d_{\min}$ and a unit cell $U=P\times R^{n-l}$, where $P$ is a parallelepiped in the $l$-dimensional subspace $R^l$ with the orthogonal subspace $R^{n-l}$ in $\R^n$. 
Consider the \emph{width} $w=\sup\limits_{u,v\in P}|\vec u-\vec v|$ and the \emph{height} $h$ equal to the maximum distance between points in the orthogonal projection of $S$ to $R^{n-l}$.
If the motif $M=S\cap U$ consists of $m$ points, then $\PDD(S;k)$ can be computed for any $k\geq 1$ in time
$$O(km(2^{4n}\log k+\log m)+2^{12n}m\log^2 k +(2^{8n}/l)k\log k+2^{8n}a^l b k),$$
where $a=1+\dfrac{2.5w+2h}{\PPC(S)}$ and $b=\log(2\PPC(S)+3w+5h)-\log d_{\min}$.
The complexity of $\AMD(S;k)$ and invariants $\PDA(S;k),\PND(S;k)$ from Definition~\ref{dfn:ADA} is the same as for $\PDD(S;k)$, because the extra computations can be done in time $O(km)$.
\end{thm}
\begin{proof}[Proof of Theorem~\ref{thm:algorithm}]
In the notations of Lemma~\ref{lem:cylinder_size}, we have integers $1\leq l\leq n$ and a basis $\vec v_1,\dots,\vec v_n$ of $\R^n$.
The first $l$ basis vectors $\vec v_1,\dots,\vec v_l$ generate the subspace $\R^l\subset\R^n$ and the lattice $\La\subset\R^l$.
Fix the origin $0\in\R^n$ at the center of the parallelepiped $U\cap R^l$.
Then any point $p\in M=S\cap U$ is covered by the closed ball $\bar B(0;r)$ for the radius $r=\sqrt{(0.5w)^2+h^2}\leq 0.5w+h$.
By Lemma~\ref{lem:distance_bounds}, all $k$ neighbors of $p$ are covered by the closed cylinder $C(0;R)$ of the radius $R=r+\sqrt{(\PPC(S)\sqrt[l]{k}+w)^2+h^2}\leq\PPC(S)\sqrt[l]{k}+1.5w+2h$.
To generate all $\La$-translates of $M$ within $C(0;R)$, we gradually extend $U$ in cylindrical layers by adding more shifted cells $U+\vec v$ for vectors $v\in\La$ until we get the upper union $U^+(0;R)$ covering the cylinder $C(0;R)$.
The upper union $U^+(0;R)$ includes $k$ neighbors of each motif point and has the size $\nu=|S\cap U^+(0;R)|=m^+(0;R)m$ estimated by Lemma~\ref{lem:cylinder_size}:
\begin{align*}
& \nu\leq\left(\dfrac{R+w}{\PPC(S)}\right)^l
\leq\left(\dfrac{\PPC(S)\sqrt[l]{k}+2.5w+2h}{\PPC(S)}\right)^l
=\left(\sqrt[l]{k}+\dfrac{2.5w+2h}{\PPC(S)}\right)^l =\\
&=k\left(1+\dfrac{2.5w+2h}{\PPC(S)\sqrt[l]{k}}\right)^l
\leq k\left(1+\dfrac{2.5w+2h}{\PPC(S)}\right)^l 
=a^l k, \text{ where } a=1+\dfrac{2.5w+2h}{\PPC(S)}.
\end{align*}

For the nearest neighbor search \cite{elkin2022counterexamples}, we can build a compressed cover tree on $\nu$ points of $T=S\cap U^+(0;R)$ in time $O(\nu c_{\min}^{8}\log\frac{2R+h}{d_{\min}})$ by \cite[Theorem~3.7]{elkin2023new}, where $c_{\min}\leq 2^n$ is the minimized expansion constant of $T$, and $\frac{2R+h}{d_{\min}}$ is the upper bound for the ratio of max/min inter-point distances.
Then $R\leq\PPC(S)\sqrt[l]{k}+1.5w+2h$ gives
\begin{align*}
& \log (2R+h)\leq \log(\sqrt[l]{k}(2\PPC(S)+3w+5h))=\log(2\PPC(S)+3w+5h)+(\log  k)/l, \\
& \text{so }
\log\dfrac{2R+h}{d_{\min}}\leq b+\dfrac{1}{l}\log  k, \text{ where } b=\log(2\PPC(S)+3w+5h)-\log d_{\min}.
\end{align*}

Using a compressed cover tree on $T$ by \cite[Theorem~4.9]{elkin2023new}, we can find $k$ neighbors of $m$ points from $S\cap U$ among $\nu$ points of $T$ in time $O\big(m c^2 (\log k) (c_{\min}^{10}\log\nu+ ck)\big)$, where $c_{\min}\leq c\leq 2^n$ are expansion constants of $T$. 
Since $\log\nu\leq\log k + l\log a$, we compute distances from each of $m$ points to their $k$ nearest neighbors in $T$ in time
\begin{align*}
& O\big(\nu c_{\min}^{8} (b+(\log  k)/l) \big) + 
O\big(m c^2 \log k (c_{\min}^{10}\log\nu+ ck)\big)\leq \\
& O\big(a^l k 2^{8n} (b+(\log  k)/l) \big) + 
O\big(m 2^{2n} \log k (2^{10n}(\log k + l\log a)+ 2^{2n} k)\big)\leq \\
& O\big( 2^{8n} a^l b k + (2^{8n}/l)k\log  k\big) + 
O\big(2^{4n} m (k\log k +2^{8n}(\log^2 k+l(\log a)\log k)) \leq \\
& O\big(2^{4n}(m+2^{4n}/l)k\log k+2^{12n}m\log^2 k+ 2^{8n}a^l b k\big), \text{ we used } l\log a\leq O(\log k).
\end{align*}

The ordered lists of distances from points $p\in S\cap U$ to their $k$ nearest neighbors in $T$ are the rows of the matrix $D(S;k)$.
If convenient, we can lexicographically sort $m$ lists of $k$ ordered distances, which needs time $O(km\log m)$, because a comparison of ordered lists of the length $k$ takes $O(k)$ time.  
The total time for $\PDD(S;k)$ is
\begin{align*}
& O(2^{4n}(m+2^{4n}/l)k\log k+2^{12n}m\log^2 k+2^{8n}a^l b k)+O(km\log m) = \\
& O(km(2^{4n}\log k+\log m)+2^{12n}m\log^2 k +(2^{8n}/l)k\log k+2^{8n}a^l b k).
\end{align*}
\end{proof}

The worst-case estimate in Theorem~\ref{thm:algorithm} is conservative due to the upper bound $2^n$ for the expansion constants $c_{\min},c$ from \cite[Definition~1.4]{elkin2023new}.
We conjecture that this upper bound can be reduced to $2^l$ for any $l$-periodic point set $S\subset\R^n$. 
\smallskip

For any fixed dimensions $l\leq n$, if we ignore the parameters $a,b,d_{\min}$, and $\PPC(S)$, then the complexity in Theorem~\ref{thm:algorithm} becomes $O(km(\log k+\log m))$, which is near-linear in both $k,m$. 
For the most practical dimensions $l=n=3$, experiments in section~\ref{sec:experiments} will report running times in minutes on a modest desktop computer. 
\smallskip

\section{Lipschitz continuous Earth Mover's Distance on invariants}
\label{sec:continuity}

This section proves the continuity of the vectorial invariants $\AMD,\ADA,\AND$, matrix invariants $\PDD,\PDA,\PND$, and their averages.
We will use the Earth Mover's Distance ($\EMD$) \cite{rubner2000earth}, which is well-defined for any weighted distributions of different sizes.
\smallskip

Definition~\ref{dfn:EMD} of $\EMD$ makes sense for any matrix invariant $I(S)$ that is an unordered collection of row vectors $\vec R_i(S)$ with weights $w_i(S)\in(0,1]$ satisfying $\sum\limits_{i=1}^{m(S)} w_i(S)=1$.
Each row $\vec R_i(S)$ should have a size independent of $i$, e.g. the number $k$ of neighbors in $\PDD(S;k)$. 
For any 
$\vec R_i=(r_{i1},\dots,r_{ik})$ and $\vec R_j=(r_{j1},\dots,r_{jk})$, the \emph{Minkowski} distance is 
$L_{q}(\vec R_i,\vec R_j)=\big(\sum\limits_{l=1}^k |r_{il}-r_{jl}|^q\big)^{1/q}$, 
$L_{\infty}(\vec R_i,\vec R_j)=\max\limits_{l=1,\dots,k}|r_{il}-r_{jl}|$.
We illustrate $\EMD$ for perturbations that scale up a unit cell, as in Fig.~\ref{fig:noise_double_triangles_peaks}~(left).
The integer sequence $\Z$ has $\PDD(\Z;2)=(1;1,1)$, a single row of weight 1 and unit distances to 2 neighbors.
The periodic sequence $\Z_\ep=\{0,1+\ep,2-\ep\}+3\Z$ is obtained from $\Z$ by $\ep$-perturbations of points $1,2$ and all their translates with period 3.
Then $\PDD(\Z_\ep;2)=\left(\begin{array}{c|cccc} 
1/3 & 1+\ep & 1+\ep \\
2/3 & 1-2\ep & 1+\ep
\end{array} \right)$, where the 2nd row represents the shifted points $1+\ep,2-\ep$.
After splitting $\PDD(\Z;2)=(1;1,1)$ into two identical rows of weights $\frac{1}{3},\frac{2}{3}$ and using $L_\infty$ on vectors of two distances, a difference between $\PDD$s can be defined as the weighted average $\frac{1}{3}\ep+\frac{2}{3}2\ep=\frac{5}{3}\ep$, which is generalized below. 

\begin{dfn}[Earth Mover's Distance $\EMD_q$] 
\label{dfn:EMD}
Let discrete sets $S,Q$ in a metric space have weighted distributions $I(S),I(Q)$ as above.
For any real $q\geq 1$, the \emph{Earth Mover's Distance}  
$\EMD_q(I(S),I(Q))=\sum\limits_{i=1}^{m(S)} \sum\limits_{j=1}^{m(Q)} f_{ij} L_q(\vec R_i(S),\vec R_j(Q))$ is minimized over variable parameters $f_{ij}\in[0,1]$
 subject to the conditions
$\sum\limits_{j=1}^{m(Q)} f_{ij}=w_i(S)$ for $i=1,\dots,m(S)$ and
$\sum\limits_{i=1}^{m(S)} f_{ij}=w_j(Q)$ for $j=1,\dots,m(Q)$. 
\end{dfn}

The first condition $\sum\limits_{j=1}^{m(Q)} f_{ij}\leq w_i(S)$ means that not more than the weight $w_i(S)$ of the vector $\vec R_i(S)$ `flows' into all vectors $\vec R_j(Q)$ via optimized parameters $f_{ij}\in[0,1]$ for $j=1,\dots,m(Q)$. 
The second condition $\sum\limits_{i=1}^{m(S)} f_{ij}=w_j(Q)$ means that all `flows' $f_{ij}$ from $\vec R_i(S)$ for $i=1,\dots,m(S)$ `flow' into $\vec R_j(Q)$ up to the maximum weight $w_j(Q)$.
\smallskip
 
The EMD satisfies all metric axioms \cite[appendix]{rubner2000earth}, needs $O(m^3\log m)$ time for distributions of a maximum size $m$, and can be approximated in $O(m)$ time \cite{shirdhonkar2008approximate}.
\smallskip

The Lipschitz continuity of invariants in $\EMD$ will use bounded perturbations of points up to $\ep$ in the metric $d_X$ of an ambient space $X$.
Since atoms are not outliers or noise, such perturbations can be formalized as the \emph{bottleneck distance} 
$d_B(S,Q)=\inf\limits_{g:S\to Q}\;\sup\limits_{p\in S}d_X(g(p),p)$ minimized over all bijections $g:S\to Q$ between (possibly infinite) sets.
This definition is computationally intractable even for finite sets due to exponentially many 
$m!$ bijections between sets of $m$ points.
\cite[Example~2.1]{widdowson2022resolving} shows that the 1-dimensional lattices $\Z$ and $(1+\de)\Z$ have $d_B=+\infty$ for any $\de>0$.
\smallskip

If $S,Q$ are lattices of equal density (equal unit cell volume), they have a finite bottleneck distance $d_B$ by \cite[Theorem~1(iii)]{duneau1991bounded}. 
If we consider only periodic point sets $S,Q\subset\R^n$ with the same density (or unit cells of the same volume), $d_B(S,Q)$ becomes a well-defined \emph{wobbling} distance \cite{carstens1999geometrical}, which is still discontinuous under perturbations by \cite[Example~2.2]{widdowson2022resolving}, see related results for non-periodic sets in \cite{laczkovich1992uniformly}.
\smallskip

Recall that the \emph{packing radius} $r(S)$ is the minimum half-distance between any points of $S$, or $r(S)$ is the maximum radius $r$ to have disjoint open balls of radius $r$ centered at all points of $S$.
Theorem~\ref{thm:continuity} extends \cite[Theorem 4.3, proved in appendix C]{widdowson2022resolving} to finite and $l$-periodic sets, and distances based on any Minkowski metric $L_q$.

\begin{thm}[Lipschitz continuity] 
\label{thm:continuity} 
Let $M$ be a finite subset of a discrete set $S$ in a space $X$ with a metric $d_X$.
Let $Q$ and its finite subset $T$ be obtained from $S$ and $M$, respectively, by perturbing every point of $S$ up to $\ep$ in the metric $d_X$.
Fix any real $q\in[1,+\infty]$ and an integer $k\geq 1$.
Interpret $\sqrt[q]{k}$ as 1 in the limit case $q=+\infty$.
\smallskip

\noindent
(a) Then $\EMD_q(\PDD(S,M;k),\PDD(Q,T;k))\leq 2\ep\sqrt[q]{k}$.
\smallskip

\noindent
(b) If $S,Q$ are $l$-periodic and 
$\min\{r(S),r(Q)\}>\ep$, then $\PPC(S)=\PPC(Q)$, and \\ 
$\EMD_q(\PDA(S;k),\PDA(Q;k))\leq 2\ep\sqrt[q]{k}$, 
$\EMD_q(\PND(S;k),\PND(Q;k))\leq 
\dfrac{2\ep\sqrt[q]{k}}{\PPC(S)}$.
\end{thm}

Theorem~\ref{thm:continuity} is proved in appendix~\ref{sec:PDD_ext_proofs} similar to \cite[Lemma 8]{widdowson2022average} for $q=+\infty$.
All columns of $\PDD,\PDA,\PND$ are ordered by the index $k$ of neighbors.
Though their rows are unordered (as points of a motif $M$), all such matrices even with different numbers of rows can be compared by Earth Mover's Distance, or by any other metrics on weighted distributions, see Definition~\ref{dfn:EMD}. 
We can simplify any $\PDD$ into a fixed-size matrix, which can be flattened into a vector, while keeping the continuity and almost all invariant data. 
Any distribution of $m$ unordered values can be reconstructed from its $m$ moments below.
When all weights $w_i$ are rational as in our case, the distribution can be expanded to equal-weighted values $a_1,\dots,a_m$.
The $m$ moments can recover all $a_1,\dots,a_m$ as roots of a degree $m$ polynomial whose coefficients are expressed via the $m$ moments \cite{macdonald1998symmetric}, e.g. any $a,b\in\R$ can be found from $a+b,a^2+b^2$ as the roots of $x^2-(a+b)x+ab$, where $ab=\frac{1}{2}((a+b)^2-(a^2+b^2))$.
\smallskip

Let $A$ be any unordered set of real numbers $a_1,\dots,a_m$ with weights $w_1,\dots,w_m$, respectively, such that $\sum\limits_{i=1}^m w_i=1$.
For any integer $t\geq 1$, the $t$-th \emph{moment} \cite[section~2.7]{keeping1995introduction} is 
$\mu_t(A)=\sqrt[t]{m^{1-t}\sum\limits_{i=1}^m w_i a_i^t}$, so
 $\mu_1(A)=\sum\limits_{i=1}^m w_i a_i$ is the usual average.
\smallskip
 
For any integer $t\geq 2$, we avoid subtracting $\mu_1$ from the numbers $a_1,\dots,a_m$, which would convert $\mu_2$ into the standard deviation $\si$, and normalize by the factor $m^{(1/t)-1}$ to guarantee the continuity of moments with the Lipschitz constant $\la=2$.

\begin{dfn}[$t$-moments matrix ${\mu^{(t)}}$]
\label{dfn:moments}
Fix any integer $t\geq 1$.
Let $I(S)$ be a matrix invariant of a cell-periodic set $S$.
For every column $A$ of $I(S)$, consisting of unordered numbers with weights, write the column $(\mu_1(A),\dots,\mu_t(A))$.
All new columns form the \emph{$t$-moments matrix} $\mu^{(t)}[I(S)]$, which has $t$ canonically ordered rows.
\end{dfn}

For $t=1$, the $1\times k$ matrix $\mu^{(1)}[\PDD(S;k)]$ appeared in Definition~\ref{dfn:PDD} as the vector $\AMD(S;k)=(\AMD_1,\dots,\AMD_k)$.
All rows and columns of the matrix $\mu^{(t)}[I(S)]$ are ordered, but this matrix is a bit weaker than $I(S)$ because each column can be reconstructed from its moments (for a large enough $t$) only up to permutation.
However, we can flatten any matrix $\mu^{(t)}[I(S)]$ 
to a vector for machine learning 
\cite{balasingham2024material,balasingham2024accelerating}.
\smallskip

Theorem~\ref{thm:lower_bound} extends \cite[Theorem~4.2, proved in appendix C]{widdowson2022resolving} to the new invariants $\PDA,\PND$ of any finite and $l$-periodic sets for a Minkowski metric $L_q$, $q\geq 1$. 

\begin{thm}[lower bounds of $\EMD$]
\label{thm:lower_bound}
For finite or $l$-periodic sets $S,Q\subset\R^n$, 
\smallskip

\noindent
(a) $\EMD_q(\PDD(S;k),\PDD(Q;k))\geq L_q(\AMD(S;k), \AMD(Q;k))$;
\smallskip

\noindent
(b) $\EMD_q(\PDA(S;k),\PDA(Q;k))\geq L_q(\ADA(S;k), \ADA(Q;k))$;
\smallskip

\noindent
(c) $\EMD_q(\PND(S;k),\PND(Q;k))\geq L_q(\AND(S;k), \AND(Q;k))$ for any $q,k\geq 1$. 
\end{thm}
\smallskip

\section{Generic completeness of Pointwise Distance Distributions}
\label{sec:generic}

We prove the generic completeness in both finite (easy) and periodic (much harder) cases.

\begin{thm}
\label{thm:gen_complete_finite}
Any cloud $C\subset\R^n$ of $m$ unordered points with distinct inter-point distances can be reconstructed from $\PDD(C;m-1)$, uniquely under isometry. 
\end{thm}
\begin{proof}[Proof of Theorem~\ref{thm:gen_complete_finite}]
Since all inter-point distances are distinct, every such distance $|p-q|$ between points $p,q\in C$ appears twice in $\PDD(C;m-1)$: once in the row of $p$ and once in the row of $q$.
After choosing an arbitrary order of points, $\PDD(C;m-1)$ suffices to reconstruct the classical distance matrix on ordered points.
This distance matrix determines $C\subset\R^n$ uniquely under isometry 
\cite{kruskal1978multidimensional}. 
\end{proof}

\begin{conj}[completeness of $\PDD$ in $\R^2$]
\label{conj:PDD_complete_n=2}
Any cloud $C\subset\R^2$ of $m$ unordered points can be reconstructed from $\PDD(C;m-1)$ uniquely under isometry. 
\end{conj}

\begin{thm}[completeness of $\PDD$ for $m\leq 4$ points]
\label{thm:PDD_complete_m<5}
$\PDD(C;m-1)$ is a complete isometry invariant of all clouds $C\subset\R^n$ for any $m\leq 4$ unordered points. 
\end{thm}

For a periodic point set $S\subset\R^n$, the generic completeness of $\PDD$ is much harder because infinitely many distances between points of $S$ are repeated due to periodicity.
We introduce a few auxiliary concepts to define \emph{distance-generic} periodic sets later.
\smallskip

For any point $p$ in a lattice $\La\subset\R^n$, the open \emph{Voronoi domain} $V(\La;p)=\{q\in\R^n \text{ such that } |q-p|<|q-p'| \mbox{ for any }p'\in\La-p\}$ is the neighborhood of all points $q\in\R^n$ that are strictly closer to $p$ than to all other points $p'$ of the lattice $\La$  \cite{smith2022practical}.
\smallskip

The Voronoi domains $V(\La;p)$ of different points $p\in\La$ are disjoint translation copies of each other and their closures tile $\R^n$, so $\cup_{p\in\La}\bar V(\La;p)=\R^n$.
For example, for a generic lattice $\La\subset\R^2$, the domain $V(\La;p)$ is a centrally symmetric hexagon.
\smallskip

Points $p,p'\in\La$ are \emph{Voronoi neighbors} if their Voronoi domains share a boundary point, so $\bar V(\La;p)\cap\bar V(\La,p')\neq\emptyset$.
Below we always assume that any lattice $\La$ is shifted to contain the origin $0$, also any periodic point set $S=\La+M$ has a point at $0$.

\begin{dfn}[neighbor set $N(\La)$ and basis distances]
\label{dfn:Nset}
For any lattice $\La\subset\R^n$, the \emph{neighbor set} of the origin 0 is $N(\La)=\La\cap\bar B(0;r)\setminus\{0\}$ for a minimum radius $r$ such that $N(\La)$ is not contained in any affine $(n-1)$-dimensional subspace of $\R^n$, and $N(\La)$ includes all $n+1$ nearest neighbors (within $\La$) of any point $q\in V(\La;0)$.
\smallskip

Consider all sets of unordered points $p_1,\dots,p_n\in N(\La)$ that are \emph{linearly independent}, i.e. the vectors $\vec p_1,\dots,\vec p_n$ form a linear basis of $\R^n$.
For any point $q\in V(\La;0)$, a lexicographically smallest list of distances $d_1(q)\leq\dots\leq d_n(q)$ from $q$ to a set of linearly independent points $p_1,\dots,p_n\in N(\La)$ is called the list of \emph{basis distances} of $q$.
\end{dfn}

The linear independence of vectors $\vec p_1,\dots,\vec p_n$ in Definition~\ref{dfn:Nset} guarantees that any point $q$ is uniquely determined in $\R^n$ by its distances $|q|,d_1(q),\dots,d_n(q)$ to $n+1$ neighbors $0,p_1,\dots,p_n$, which are not in the same $(n-1)$-dimensional subspace.
\smallskip

Let $\La$ be generated by $(2,0),(0,1)$.
The Voronoi domain $V(\La;0)$ is the rectangle $(-1,1)\times(-0.5,0,5)$.
The neighbor set $N(\La)\subset\La$ includes the 3rd neighbors $(0,\pm 2)$ of the points $(0,\pm 0.4)\in V(\La;0)$.
Indeed, if in Definition~\ref{dfn:Nset} $\La$ has a radius $r<2$, then $\La\cap\bar B(0;r)\setminus\{0\}=\{(0,\pm 1)\}$ is in the 1-dimensional subspace ($y$-axis) of $\R^2$.
For $q=(0,0.4)$, considering all pairs $(\vec p_1,\vec p_2)$ that generate $\R^2$ among the four possibilities $((0,\pm 1),(\pm 2,0))$, we find the basis distances $d_1(q)=0.6<d_2(q)=\sqrt{0.4^2+2^2}\approx 2.04$ for the 2nd and 3rd lattice neighbors $p_1=(0,1)$ and $p_2=(\pm 2,0)$ of $q$, respectively.

\begin{lem}
\label{lem:Nset_bounds} 
The neighbor set $N(\La)$ of any lattice $\La$ is covered by $\bar B(0;2R(\La))$, where the \emph{covering radius} $R(\La)$ is the minimum $R>0$ such that $\cup_{p\in\La}\bar B(p;R)=\R^n$.
\end{lem}
\begin{proof}[Proof of Lemma~\ref{lem:Nset_bounds}]
Any point $p$ in the closure $\bar V(\La;0)$ of the Voronoi domain has $n+1$ lattice neighbors (within $\La$) among them the origin $0\in\La$ and at least $2(2^n-1)$ Voronoi neighbors of $0$ \cite{conway1992low}. 
In $\R^n$, any vertex of the boundary of $V(\La;0)$ is equidistant to at least $n+1$ points of $\La$ (the origin $0$ and its $n$ Voronoi neighbors).
The longest of these distances to Voronoi neighbors is the covering radius $R(\La)$.
The ball $\bar B(0;2R(\La))$ covers all Voronoi neighbors of $0$ and hence the neighbor set $N(\La)$.
\end{proof}

\begin{dfn}[a distance-generic set]
\label{dfn:generic_set}
A periodic point set $S=M+\La\subset\R^n$ with the origin
$0\in\La\subset S$ is called \emph{distance-generic} if the following conditions hold.
\smallskip

\noindent
(\ref{dfn:generic_set}a)
For any points $p,q\in S\cap V(\La;0)$, the vectors $\vec p,\vec q$ are not orthogonal.
\smallskip

\noindent
(\ref{dfn:generic_set}b)
For vectors $\vec u,\vec v$ between any two pairs of points in $S$, if $|\vec u|=l|\vec v|\leq 2R(\La)$ for $l=1,2$, then $\vec u=\pm l\vec v$ and $\vec v\in\La$.
\smallskip

\noindent
(\ref{dfn:generic_set}c)
For any point $q\in S\cap V(\La;0)$, let $d_0=|q|$ be its distance to the closest neighbor $p_0=0$ in $\La$.
Take any linearly independent points $p_1,\dots,p_n\in N(\La)$ and any distances $d_1\leq\dots\leq d_n$ from $q$ to some points in $S\cap \bar B(0;2R(\La))$.
The $n+1$ spheres $\bd B(p_i;d_i)$ can meet at a single point of $S\cap V(\La;0)$ only if $d_1\leq\dots\leq d_n$ are the basis distances of $q$ and only for two tuples $p_1,\dots,p_n\in N(\La)$ related by 
$\vec v\mapsto -\vec v$.
\end{dfn}

Condition~(\ref{dfn:generic_set}b) means that all inter-point distances are distinct apart from necessary exceptions due to periodicity.
Since any periodic set $S=M+\La\subset\R^n$ is invariant under translations along all vectors of $\La$, condition~(\ref{dfn:generic_set}b) for $|\vec v|\leq 2R(\La)$ can be checked only for vectors from all points of $S$ in the original Voronoi domain $V(\La;0)$ to all points in the domain $3V(\La;0)$ extended by factor 3. 
Condition~(\ref{dfn:generic_set}b) implies that $S$ has no points on the boundary $\bd V(\La;0)$, because any such point is equidistant to points $0,v\in\La$ and hence should belong to $\La$. 
Let a \emph{lattice distance} be the Euclidean distance from any $p\in M=S\cap V(\La;0)$ to its lattice translate $p+\vec v$ for all $\vec v\in\La$. 
Condition~(\ref{dfn:generic_set}a) guarantees that only a lattice distance $d$ appears together with $2d$ (and possibly with higher multiples) in a row of $\PDD(S;k)$.
Any such $d$ and its multiples are repeated twice in every row, because $\La$ is centrally symmetric. 

\begin{lem}[almost any periodic set is distance-generic]
\label{lem:distance-generic}
Let $S=M+\La\subset\R^n$ be any periodic point set. 
For any $\ep>0$, one can perturb coordinates of a basis of $\La$ and of points from $M$ up to $\ep$ such that the resulting perturbation $S'$ of $S$ is a distance-generic periodic point set in the sense of Definition~\ref{dfn:generic_set}. 
\end{lem}
\begin{proof}
We can assume that the motif $M$ of $S$ is a subset of the open Voronoi domain $V(\La;0)$ and include the origin $0$.
We show below that conditions (\ref{dfn:generic_set}a,b) define a codimension 1 \emph{discriminant} (singular subspace) in the space of all parameters $P$ that are coordinates of points of $M$ and of basis vectors of $\La$.
In condition~(\ref{dfn:generic_set}a), for any points $p,q\in V(\La;0)$, the orthogonality is expressed as $f_a(p,q)=\vec p\cdot\vec q = \sum\limits_{i=1}^n p_i q_i=0$.  
In condition~(\ref{dfn:generic_set}b), for any vectors $\vec u,\vec v$ that join points of $S$, have a maximum length $2R(\La)$, and satisfy $u\neq \pm l\vec v$ for $l=1,2$, the equality $|\vec u|= l|\vec v|$ can be written as $f_b(u,v)=\sum\limits_{i=1}^n u_i^2-l^2 \sum\limits_{i=1}^n v_i^2=0$.
So condition~(\ref{dfn:generic_set}a) forbids a codimension 1 subspace defined by finitely many equations $f_b(u,v)=0$ for all $u,v$ above.  
\smallskip

Similarly, condition (\ref{dfn:generic_set}c) can be written via polynomial equations in point coordinates.
For any fixed radii $d_0,\dots,d_n$, almost all $n+1$ spheres in $\R^n$, whose centers are not in any $(n-1)$-dimensional affine subspace, have no common points.
Hence condition (\ref{dfn:generic_set}c) also forbids a codimension 1 subspace.
All involved functions in equations above are continuous in the coordinates of points and basis vectors.
Then a motif $M=S\cap V(\La;0)$ and a basis of $\La$ can be slightly perturbed to move $S$ to $S'$ outside the union of all finitely many codimension 1 subspaces above.
Hence any periodic point set $S$ can be made distance-generic by a small enough perturbation. 
\end{proof}
\smallskip

The size $m$ of a motif $M$ is an isometry invariant because any isometry maps $N$ to another hose motif of the same size.
In dimensions $n=2,3$, any lattice $\La$ can be reconstructed from its complete isometry invariants \cite{kurlin2024mathematics,kurlin2022complete}.
Theorem~\ref{thm:gen_complete} reconstructs a periodic point set $S=M+\La\subset\R^n$ in any dimension $n\geq 2$ from the invariant $I(S)$ consisting of $m$, $\PDD(S;k)$, and (complete invariants of) a lattice $\La$ to satisfy completeness (\ref{pro:map}a) for distance-generic periodic sets $S\subset\R^n$. 
New Lemma~\ref{lem:distance-generic} and the arguments below clarify the early unreviewed proof in \cite[appendix C]{widdowson2022resolving}.
 
\begin{thm}[generic completeness of $\PDD$]
\label{thm:gen_complete}
Let $S=M+\La\subset\R^n$ be any distance-generic periodic set whose motif $ M$ has $m$ points.
Let $R(\La)$ be the smallest radius $R$ such that all closed balls with centers $p\in\La$ and radius $R$ cover $\R^n$.
For any $k$ such that all distances in the last column of $\PDD(S;k)$ are larger than $2R(\La)$, $S$ can be reconstructed from $m$, $\La$, and $\PDD(S;k)$, uniquely under isometry in $\R^n$.
\end{thm}
\begin{proof}
The given number $m$ of points in a unit cell $U$ of $\La$ is
a common multiple of all denominators in rational weights of the rows in the given matrix $\PDD(S;k)$.
Enlarge $\PDD(S;k)$ by replacing every row of a weight $w$ with the integer number $mw$ of identical rows having the same weight $\frac{1}{m}$.
One can assume that the origin $0\in\La$ belongs to the motif $M$ of $S$ and is represented by the first row of $\PDD(S;k)$.
\smallskip

If $\PDD(S;k)$ has $m\geq 2$ rows, we will reconstruct all other $m-1$ points of the periodic point set $S$ within the open Voronoi domain $V(\La;0)$.
No points of $S$ can be on the boundary of $V(\La;0)$ due to condition~(\ref{dfn:generic_set}b) on distinct distances. 
\smallskip

Remove from each row of $\PDD(S;k)$ all \emph{lattice distances} between any points of $\La$.
Then every remaining distance is between only points $p,q\in S$ such that $\vec p-\vec q\not\in\La$.
Take a unique point $q\in S\cap V(\La;0)\setminus\{0\}$ that has the smallest distance $d_0=|q|$ to the origin and hence uniquely determined in the row of $q$ in $\PDD(S;k)$.
Then we will look for $n$ basis distances $d_1<\dots<d_n$ from $q$ to its further $n$ lattice neighbors $p_1,\dots,p_n\in N(\La)\subset\La-0$ such that $\vec p_1,\dots,\vec p_n$ form a linear basis of $\R^n$.
All basis distances $d_0,\dots,d_n$ are distinct due to (\ref{dfn:generic_set}b).
By Lemma~\ref{lem:Nset_bounds} they appear once in both rows of the points $0,q\in S$ in $\PDD(S;k)$ after 
the shortest distance $d_0=|q|$. 
\smallskip

Though the basis distances of $q$ may not be the $n$ smallest values appearing after $d_0=|q|$ in the first and second rows of $\PDD(S;k)$, we will try all subsequences $d_1<\dots<d_n$ of distinct distances shared by both rows.
Similarly, we cannot be sure that $n$ closest neighbors of $q$ in $S\setminus\{0\}$ define linearly independent vectors of $\La$.
\smallskip

Hence we try all linearly independent points $p_1,\dots,p_n\in N(\La)$.
For all finitely many choices, we check if the $n+1$ spheres $\bd B(p_i;d_i)$ meet at a single point in $V(\La;0)$, which will be the required point $q$.
These $(n-1)$-dimensional spheres are 1D circles for $n=2$ and 2D spheres for $n=3$.
Condition~(\ref{dfn:generic_set}c) will guarantee below a reconstruction of $q$ as a single intersection of these $n+1$ spheres of dimension $n-1$. 
\smallskip

The basis distances $d_1<\dots<d_n$ of $q$ should form the lexicographically smallest list among all lists of distances from $q$ to points $p_1,\dots,p_n\in N(\La)$. 
This smallest list emerges for at most two tuples of linearly independent points $p_1,\dots,p_n\in N(\La)$ related by the isometry $\vec v\mapsto-\vec v$, which preserves $\La$. 
For a first reconstruction outside $\La$, we choose any of these tuples and find the intersection point $q=\cap_{i=0}^n \bd B(p_i;d_i)$. 
\smallskip

Any other point $p\in (S\setminus\{0,q\})\cap V(\La;0)$ is uniquely determined similarly to the point $q$ above by using its basis distances $d_0(p)<d_1(p)<\dots<d_n(p)$ to points $0=p_0,p_1,\dots,p_n\in N(\La)$.
At the end of reconstruction, we have a final choice between $\pm p$ symmetric with respect to the origin $0$.
Since the second point $q$ is already fixed, the third point $p$ is also restricted by the distance $|p-q|$ appearing once only in the second and third rows of $\PDD(S;k)$.
The distance $|p-q|$ doesn't help to resolve the ambiguity between $\pm p$ only if $q$ belongs to the bisector of points equidistant to $\pm p$.
In this case, $p,0,q$ form a right-angle triangle, which is forbidden by condition~(\ref{dfn:generic_set}a).
Hence $p$ 
is uniquely determined by the already fixed point $q$ and lattice $\La$.
\end{proof}
\smallskip

\section{Detecting near-duplicates in the world's largest databases}
\label{sec:experiments}

This section reports thousands of previously unknown (near-)duplicates in the world's largest databases \cite{taylor2019million,gravzulis2009crystallography,zagorac2019recent,jain2013commentary}. 
The sizes in Table~\ref{tab:databases} below are the numbers of all periodic crystals (with no disorder and full geometric data) in September 2024 (total number is 1,847,462, see Table~SM6 and all experimental details in appendix~\ref{sec:exp_details}. 
\smallskip

\begin{table}[h!]
\caption{Links and versions of the world's largest materials databases, see their sizes in Table~SM6.}
\label{tab:databases}
\begin{center}
\begin{tabular}{ll}
database and web address & version \\
\hline
CSD: Cambridge Structural Database,  http://ccdc.cam.ac.uk & version 6.00  \\
COD: Crystallography Open Database, crystallography.net/cod & July 30, 2024 \\
ICSD: Inorganic Crystal Structures, icsd.products.fiz-karlsruhe.de & Feb 25, 2025 \\
MP: Materials Project, http://next-gen.materialsproject.org & v2023.11.1 \\ 
GNoME: github.com/google-deepmind/materials\_discovery & Nov 29, 2023 \\
\end{tabular}
\end{center}
\end{table}

We first used the vector $\ADA(S;100)$ to find nearest neighbors across all databases by $k$-d trees \cite{gieseke2014buffer} up to $L_\infty\leq 0.01\angstrom$.
Since the smallest inter-atomic distances are about $1\angstrom=10^{-10}$m, atomic displacements up to $0.01\angstrom$ are 
considered experimental noise.
For the closest pairs found by $\ADA(S;100)$, the stronger $\PDA(S;100)$
can have only equal or larger $\EMD\geq L_\infty$ by Theorem~\ref{thm:lower_bound}.
The CSD, COD, ICSD should contain experimental structures.
MP is obtained from ICSD by extra optimization.
\smallskip

Table~\ref{tab:EMD_Linf_PDA100_leq001A} shows that the well-curated 60-year-old CSD has 0.9\% near-duplicate crystals, while more than a third of the ICSD consists of near-duplicates that are geometrically almost identical so that all atoms can be matched by an average perturbation up to $0.01\angstrom$.
Table~1 in \cite[section~6]{anosova2024importance} reported many thousands of exact duplicates, where chemical elements were replaced while keeping all coordinates fixed.  
These replacements are physically impossible without more substantial perturbations. Five journals are investigating integrity \cite{chawla2023crystallography}, see details in appendix~\ref{sec:exp_details}.
\smallskip

The bold numbers in Table~\ref{tab:EMD_Linf_PDA100_leq001A} count near-duplicates and their percentages within each database, which should be filtered out, else the ground truth data becomes skewed.
Table~\ref{tab:near-duplicates_diff_cells} confirms that cell-based comparisons miss near-duplicates as in Fig.~\ref{fig:noise_double_triangles_peaks}.

\begin{table}[h!]
\begin{center}
\setlength{\tabcolsep}{3pt}
\caption{Count and percentage of all ideal periodic crystals in each database (left) found to have a near-duplicate in other databases (top) by the distance $\EMD\leq 0.01\angstrom$ on matrices $\PDA(S;100)$.}
\label{tab:EMD_Linf_PDA100_leq001A}
\begin{tabular}{l|cc|cc|cc|cc|cc}
duplicates & \multicolumn{2}{c|}{CSD} & \multicolumn{2}{c|}{COD} & \multicolumn{2}{c|}{ICSD} & \multicolumn{2}{c|}{MP} & \multicolumn{2}{c}{GNoME}  \\ 
in databases    & count  & \%     & count  & \%     & count & \%     & count & \%   & count & \%     \\ 
      \hline
CSD   & 8343 & 0.92 & 283000 & 31.19 & 26506 & 2.92 & 33 & 0.00 & 1 & 0.00\\
COD   & 286663 & 80.18 & 19568 & 5.47 & 47065 & 13.16 & 5231 & 1.46 & 2705 & 0.76\\
ICSD  & 26853 & 15.78 & 69948 & 41.10 & 51085 & 30.01 & 27194 & 15.98 & 15449 & 9.08\\
MP    & 73 & 0.05 & 11986 & 7.82 & 15188 & 9.91 & 19177 & 12.51 & 10681 & 6.97\\
GNoME & 2 & 0.00 & 1800 & 0.47 & 2614 & 0.68 & 3401 & 0.88 & 82859 & 21.53
\end{tabular}
\end{center}
\end{table}

\vspace*{-2mm}
\begin{table}[h!]
\setlength{\tabcolsep}{3pt}
\begin{center}
\caption{Near-duplicates from Table~\ref{tab:EMD_Linf_PDA100_leq001A} whose unit cells differ by at least the same threshold of $0.01\angstrom$.
Unit cells are compared by $L_\infty$ between vectors of corresponding lengths of 3 edges and 3 face diagonals.}
\label{tab:near-duplicates_diff_cells}
\begin{tabular}{l|cc|cc|cc|cc|cc}
duplicates & \multicolumn{2}{c|}{CSD} & \multicolumn{2}{c|}{COD} & \multicolumn{2}{c|}{ICSD} & \multicolumn{2}{c|}{MP} & \multicolumn{2}{c}{GNoME}\\ 
in databases    & count  & \%     & count  & \%     & count & \%     & count & \%  & count & \%   \\ 
      \hline
CSD   & 776 & 0.09 & 419 & 0.05 & 210 & 0.02 & 29 & 0.00 & 1 & 0.00\\
COD   & 472 & 0.13 & 7263 & 2.03 & 8629 & 2.41 & 5059 & 1.42 & 2684 & 0.75\\
ICSD  & 462 & 0.27 & 28863 & 16.96 & 42946 & 25.23 & 26554 & 15.60 & 15360 & 9.02\\
MP    & 70 & 0.05 & 11790 & 7.69 & 14915 & 9.73 & 18582 & 12.13 & 10608 & 6.92\\
GNoME & 2 & 0.00 & 1786 & 0.46 & 2590 & 0.67 & 3346 & 0.87 & 60248 & 15.65
\end{tabular} 
\end{center}
\end{table}

In the past, the (near-)duplicates were impossible to detect at scale, because the traditional comparison through iterative alignment of 15 (by default) molecules by the COMPACK algorithm \cite{chisholm2005compack} is too slow for all-vs-all comparisons.
Tables~\ref{tab:times_duplicates_PDA100} and SM6 
compare the running times: \textbf{minutes} of $\PDA(S;100)$ vs \textbf{years} of RMSD, extrapolated for the same machine from the median time 117 milliseconds (582 ms on average) for 500 random pairs 
in the CSD. 
On the same 500 pairs, $\PDA(S;100)$ for two crystals and $\EMD$ together took only 7.48 ms on average.
All experiments were done on a typical desktop computer (AMD Ryzen 5 5600X 6-core, 32GB RAM). 

\begin{table}[h!]
\setlength{\tabcolsep}{4pt}
\caption{Running times to compute $\PDA(S;100)$ and find all near-duplicates in Table~\ref{tab:EMD_Linf_PDA100_leq001A} with $\EMD\leq 0.01\angstrom$ across all major databases (seconds in the last 4 columns), compare with years in Table~SM6.} 
\label{tab:times_duplicates_PDA100}
\begin{center}
\begin{tabular}{lrrrrrrr}
database & $\PDA$, min:sec & $\EMD$, min:sec & CSD  & COD  & ICSD  & MP & GNoME  \\ 
\hline
CSD & 60:44  & 12:21  & 125.5 & 498.1 & 77.0 & 19.1 & 20.6 \\
COD & 30:16 & 16:29 & 524.5 & 122.0  & 235.1 & 79.6 & 27.0 \\
ICSD & 5:57  & 22:04 &  80.5  & 239.3  & 515.8  & 414.9 & 73.5 \\
MP & 1:40  & 13:31 & 28.2   & 82.9 & 413.8  & 222.8 & 63.0 \\
GNoME & 4:07  & 18:59 & 29.0   & 26.7 & 74.5  & 64.5 & 943.7
\end{tabular}
\end{center}
\end{table}

\section{Discussion}
\label{sec:discussion}

For hundreds of years, crystals were classified almost exclusively by discrete tools such as space groups or by using reduced cells, which are unique in theory.
Fig.~\ref{fig:noise_double_triangles_peaks}~(left) showed that any known crystal can be disguised by changing a unit cell, shifting atoms a bit, changing chemical elements, then claimed as `new', see appendix~\ref{sec:exp_details}.
Such artificially generated structures threaten the integrity of experimental databases \cite{chawla2023crystallography}, which are skewed by previously undetectable near-duplicates.
These challenges motivated the stronger questions ``how much different?'' and ``can we get a structure from its code?'', which were formalized in Problem~\ref{pro:map} aiming for a continuous parametrization of the space of crystals.
One limitation is that $\PDD$ is not proved to be complete and a random PDD may not be realizable by a crystal because inter-atomic distances cannot be arbitrary, which we plan to improve in future work for a full solution of Problem~\ref{pro:map} in the periodic case.
However, these invariants already parametrize the `universe' containing all known crystals as `shiny stars' and all not yet discovered crystals hidden in empty spots on the same map.
Appendix~\ref{sec:exp_details} shows these geographic-style maps of all four databases in our invariant coordinates. 
\smallskip

The impact is the efficient barrier for noisy duplicates of known structures because the invariants quickly find nearest neighbors of newly claimed materials in the existing databases, as shown for all crystals from the GNoME \cite{anosova2024importance} and A-lab \cite{widdowson2025geographic} datasets. 

\bibliographystyle{siamplain}
\bibliography{SIAP_PDD_arxiv2025Dec}
\bigskip

\appendix

\section{Details of experiments on the world's largest databases} 
\label{sec:exp_details}

This appendix describes the main experiments in more detail. 
Some entries in the CSD and COD are incomplete or disordered (not periodic). After removing such entries, we were left with 852,890 CSD structures and 351,380 COD structures.
\smallskip

First we computed $\mu^{(10)}[\PDD(S;100)]$ for all entries, taking 27 min 33 sec for the CSD and 12 mins 15 sec for COD (2 ms per structure on average). 
To find exact geometric matches between databases, we use the $k$-d tree data structure, designed for fast nearest neighbor lookup. 
A $k$-d tree can be constructed from any collection of vectors, which can then be queried for a number of nearest neighbors of a new vector, using a binary tree style algorithm with logarithmic search time.
\smallskip

Then we flattened each matrix $\mu^{(10)}[\PDD(S;100)]$ to a vector with 1000 dimensions, constructed a $k$-d tree for both CSD and COD, then queried the 10 nearest neighbors for each item in the other. 
If the most distant neighbor for any entry is closer than the threshold $10^{-10}\angstrom$ (within floating point error), we extend the search and find more neighbors until all pairs within the threshold are found. 
We found a total of 278,236 geometric matches (almost exact duplicates at the atomic level); an overlap between the databases of one third of the CSD and over 80\% of the COD.
\smallskip

Of particular interest are the 235 pairs with near-zero distance but different chemical compositions.
Indeed, the impossibility of complex organic structures sharing the exact same geometry but not composition implies an error or labeling issue. 
All the pairs were confirmed as geometric duplicates by manually checking their CIFs and found to have different compositions, mostly for the three reasons given below. The 5 remaining pairs not in these three categories are in Table~\ref{tab:CSD-COD-errors} below.
\smallskip

\begin{itemize}
	\item The source CIF has atoms whose types are labelled differently by the tags `\_atom\_site\_label' and `\_atom\_site\_type\_symbol'. COD entries always use the data in the uploaded CIF, but CSD entries occasionally have data corrected and if so often have a remark describing the correction (109 pairs, Table~\ref{tab:CSD-COD-match-mixed-cif-types}).
    \item Disorder was modeled as a `mixed site' with one atomic type present and a remark on the CSD entry explaining the disorder (20 pairs, Table~\ref{tab:CSD-COD-match-mixed-sites}).
	\item Types in the CIF are consistent but CSD curators discovered incorrectly labelled atoms which were corrected and given a remark (78 pairs, Table~\ref{tab:CSD-COD-match-CSD-remark}).
\end{itemize}
\smallskip

\begin{table}
	\centering
	\begin{tabular}{ll}
		CSD ID   & COD ID  \\ \hline
		ABAGUG   & 4112689 \\
		AFUXEG   & 2238369 \\
		AJAREI   & 7113511 \\
		AJAREI   & 7103824 \\
		BAKXUH   & 8100721 \\
		BIGNUA   & 5000340 \\
		BOQBAK   & 2009202 \\
		CABSAA   & 2200584 \\
		CALWIW   & 4114997 \\
		CAQFUV   & 7027367 \\
		CUDJAP   & 1557108 \\
		DECJUS   & 4065161 \\
		DECTAI   & 4065524 \\
		DEGFOL   & 2208310 \\
		DEHKUX   & 7101047 \\
		DOBBIF   & 7213201 \\
		DUDZOS   & 4302088 \\
		EBASIN01 & 7708085 \\
		EFESUE   & 4107864 \\
		EGELUY   & 4108535 \\
		ELOJOE   & 4314231 \\
		ENIZEH   & 2018012 \\
		ESADAD   & 4062269 \\
		EVEMIB   & 4020894 \\
		EXATEC   & 7050257 \\
		EXATIG   & 7050258 \\
		FONGAQ01 & 2005101 \\
		FUPJIJ   & 7212965 \\ 
            GESJIY & 4333010 \\
		GETSAD & 7245388 \\
		GUHYOX & 7010289 \\
		HABTAF & 2001740 \\
		HIXWEQ & 2008462 \\
		IKOSIL & 4065905 \\
            JECBID & 7006569 \\
		JUCJOJ & 4003435 \\
		KABHOL & 4113866 \\
	\end{tabular}
	\hspace{1mm}
	\begin{tabular}{ll} 
		CSD ID & COD ID  \\ \hline
		KAVYOW & 7008840 \\
		KEBQUF & 7018464 \\
		KEZLOS & 4117778 \\
		KEZMUZ & 4117772 \\
		KIZFOR & 7232188 \\
		KIZJOT & 4029575 \\
		LABSAI & 2001822 \\
		LAMQEV & 4116446 \\
		LAVFAP & 2001334 \\
		LAZWOY & 2009422 \\
		LINLOJ & 2003397 \\
		LINLUP & 2003398 \\
		LUNDIH & 1507498 \\
		MEHCEI & 2208583 \\
		MEJRAV & 4101504 \\
		MENCAJ & 7009977 \\
		METSAF & 7702634 \\
		NAJQUK & 4323901 \\ 
            NEDXID01 & 2105611 \\
		NIQJIJ01 & 1549188 \\
		NOCXIM   & 4322709 \\
		NOVHUB02 & 2103787 \\
		NUMWOH   & 2007448 \\
		NUVZOV   & 4501471 \\
		ODEBII   & 4115837 \\
		OGOLUR   & 5000295 \\
		OHEFAI   & 7012100 \\
		OHEJIU   & 7204467 \\
		PAMWIK   & 2205526 \\
            PAXKEG   & 2235126 \\
		PAYSUF   & 2235091 \\
		PHOXBZ01 & 2017696 \\
		PIHJUL   & 4030494 \\
		QAHFOV   & 7012335 \\
		QAZTEQ   & 4077596 \\
		QEJYUA   & 4508631 \\
	\end{tabular}
	\hspace{1mm}
	\begin{tabular}{ll} 
		CSD ID   & COD ID  \\ \hline
		QIQNIN   & 4077174 \\
		QOQFOT   & 4348248 \\
		QUXBAN   & 2017697 \\
		RAKMOF   & 7114739 \\
		RARFUM   & 4327332 \\
		RIVKOW01 & 4310386 \\
		ROCJUP   & 4304894 \\
		RORGUA   & 4323669 \\
		RORGUA02 & 4323669 \\ 
            RUVFET & 4323710 \\
		SAQHIC & 1100776 \\
		SAQQUX & 4308912 \\
		SAXCUP & 2007898 \\
		TIPYOG & 2005914 \\
		TOCNOO & 4323981 \\
		UJECOB & 7012760 \\
		UJIKAZ & 7213431 \\
		UVOHIY & 7040448 \\
		WASKAC & 2001382 \\
		WATMIO & 4309447 \\
		WIKRIS & 8102105 \\
		WIRJEM & 2005120 \\
		XAFDUD & 4321242 \\
		XAGJUK & 8101251 \\
		XAVDEF & 4103386 \\
		XIHVOZ & 4317724 \\
		XIJNOT & 4115818 \\
		XOFXIZ & 1507458 \\
		XOFXOF & 1507459 \\
		XOPNAT & 7218637 \\
		XUFLUH & 7034643 \\
		YEJQAF & 2012123 \\
		ZAGCUJ & 1559337 \\
		ZAYRUM & 2003941 \\
		ZEXQUO & 2004127 \\
		ZIKMAH & 2004275 \\
	\end{tabular}
	\caption{109 exact geometric matches (within $10^{-10}\angstrom$) between the CSD and COD where the original CIF has atoms labelled as different types by `\_atom\_site\_label' and `\_atom\_site\_type\_symbol'. Several of the CSD entries have a remark noting that atoms were corrected in curation.}
	\label{tab:CSD-COD-match-mixed-cif-types}
\end{table}

\begin{table}
	\centering
	\begin{tabular}{ll} 
		CSD ID   & COD ID  \\ \hline
		AFUKIX & 7211182 \\
		AJIRAM01 & 2100097 \\
		BAPLOT09 & 7121265 \\
		BASLAJ & 7050473 \\
		BASMAK & 7050478 \\
		BEPWUQ & 4507409 \\
		BIHVUL & 7210243 \\
		BIKJEN & 7231097 \\
		BODZEB & 7215818 \\
		BOMMEX & 4124237 \\
		CIPDIQ & 7213596 \\
		COLNUP & 4034420 \\
		COTNAC & 7219615 \\
		DAGRUB & 4349194 \\
		DENBAD & 7710591 \\
		DIBGAX & 7151087 \\
		DISNAW & 1543965 \\
		DOJFEQ & 7230639 \\
		DOSSOW & 7123961 \\
		DUFXOS & 7104457 \\
		EMUMEF & 1503106 \\
		ETEPIC & 2203286 \\
		EWABIO & 4324780 \\
		FEBBOH & 7130024 \\
		FOBXAY & 7122779 \\
            FORWOA & 7116555 \\
		GACZEQ & 7151378 \\
		IQAFEN & 7225754 \\
		ISORIU & 7242793 \\
		ISUFAE & 7205743 \\
		JEMLAP & 4101489 \\
		JOHXUB & 7114582 \\
		KUKQIS & 7234247 \\
	\end{tabular}
	\hspace{1mm}
	\begin{tabular}{ll}
		CSD ID   & COD ID  \\ \hline
		KUTWUU & 7126770 \\
		LEBTET & 7110143 \\
		MARSIH & 4321045 \\
		MAVZUG & 7107511 \\
		MIPNEG & 4335723 \\
		MOGHAU & 7123768 \\
		NAFTIA & 7223916 \\
		NAJMER & 7050031 \\
		NEHFUE & 4131268 \\
		NEYJIM & 7021415 \\
		NIFJAO & 4022923 \\
		NIMXOY & 4334458 \\
		NUKCAZ & 7035092 \\
		NUQVAY & 7118051 \\
		OKUJOV & 4347519 \\
		OMIJIF & 7118994 \\
            PAQCEQ & 4061419 \\
		PECRUL & 4300654 \\
		PIBTAW & 1505325 \\
		PICFIR & 4072624 \\
		PIGJEW & 4080504 \\
		PINHUP & 1558382 \\
		PUTCOY & 7055058 \\
		QAMKAU & 7705818 \\
		QANLIE & 7061176 \\
		QOTVUS & 7221578 \\
		QOWKOE & 4341138 \\
            QUCXAP & 7117360 \\
		RADBAB & 7025360 \\
		REGVII & 4116980 \\
		REMVOU & 2006347 \\
		REYRES & 4116989 \\
	\end{tabular}
	\hspace{1mm}
	\begin{tabular}{ll}
		CSD ID   & COD ID  \\ \hline
            RIDYAI & 7131471 \\
		ROBKID01 & 1520266 \\
		SALGUK & 7155485 \\
		SELHAU & 4027023 \\
		SIJBAQ & 7109679 \\
		SOVZOT & 4063498 \\
		TAVWEW & 7129345 \\
		TEMMOQ & 7056766 \\
            TETQUI & 7711227 \\
		UCACAF & 7119310 \\
		UGOVER & 4115188 \\
		UGUBIJ & 7220063 \\
		UGUSIB & 1551384 \\
		UKAXUB & 7234657 \\
		UMESIQ & 7225104 \\
		UVOHOE & 7040449 \\
		UYEBUX & 7236357 \\
		VEFLUR & 1561274 \\
		VENJIJ & 4331164 \\
		VOCNUY & 7239443 \\
		WOTMEA & 4036052 \\
		WULGIV & 4036188 \\
		XEXCOV & 7045895 \\
		YEJNOU & 7710456 \\
		YEPSUI & 8000091 \\
		YURCEV & 7036965 \\
		YUYDAZ & 7037146 \\
		ZEYKIA & 7230274 \\
		ZIDBOF & 7210579 \\
		ZIGDIG & 7246585 \\
		ZUNNUU & 7059654 \\
		ZITXUV & 2004330 \\
	\end{tabular}
	\caption{97 exact geometric matches (within floating point error of $10^{-10}\angstrom$) between the CSD and COD with different chemical compositions where erroneously labelled atoms were corrected by the CSD entry in curation. Most entries have a remark mentioning the correction.}
	\label{tab:CSD-COD-match-CSD-remark}
\end{table}

\begin{table}
	\centering
	\begin{tabular}{ll}
		CSD ID   & COD ID  \\ \hline
		FIQDUI & 7713232 \\
		GODSEY & 4305065 \\
		GOHPAU01 & 2102515 \\
		LIJXAD & 2010401 \\
		LIJXEH & 2010402 \\
            MUMXIB01 & 2102385 \\
		NUTZIN & 7036504 \\
	\end{tabular}
	\hspace{1mm}
	\begin{tabular}{ll}
		CSD ID   & COD ID  \\ \hline
		NUTZOT & 7036505 \\
		QALLUL & 4505437 \\
		TIPSAM & 2101647 \\
            TIPSAM01 & 2101646 \\
		TIPSAM02 & 2101648 \\
		TOGVOA & 2005985 \\
		ZUGVIG & 2004797 \\
	\end{tabular}
	\hspace{1mm}
	\begin{tabular}{ll}
		CSD ID   & COD ID  \\ \hline
		ZUGVOM & 2004798 \\
            ZUGVUS & 2004799 \\
		ZUGWAZ & 2004800 \\
		ZUGWED & 2004801 \\
		ZUGWIH & 2004802 \\
		ZUHCOW & 2004740 \\
	\end{tabular}
	\caption{20 exact geometric matches (within floating point error $10^{-10}\angstrom$) between the CSD and COD with different compositions where disorder was modeled as a `mixed site' with only one of two atomic types listed. Usually the CSD entry has a remark describing the disorder.}
	\label{tab:CSD-COD-match-mixed-sites}
\end{table}

\begin{table}
	\centering
	\begin{tabular}{lll} 
		CSD ID & COD ID  & Remark  \\ \hline
		APEJUD & 1544509 & APEJUD has atom label `Unknown1?' \\
		HIWHEA & 4321802 & C1 $\leftrightarrow$ N1C \\ 
		IPOQOU & 4063641 & N2 $\leftrightarrow$ C22 \\ 
		LEFYIF & 4300748 & B1, B2, C5, C1 $\leftrightarrow$ C27, C17, B21, B11 \\ 
		NIDPIB & 7208250 & N2 $\leftrightarrow$ O21  
	\end{tabular}
	\caption{5 exact geometric matches (within $10^{-10}\angstrom$) between the CSD and COD with different compositions. It could not be confirmed if the last four pairs are erroneous or corrected by the CSD.}
	\label{tab:CSD-COD-errors}
\end{table}

In addition to cross-comparing the CSD and COD, we also analyzed the ICSD and Materials Project database (MP) and compared them all pairwise, as well as searching for duplicates within each database. 
Table~\ref{tab:cross_db_matches} below shows how many matches were found, and how many also shared the same composition.
\smallskip

\begin{table}
    \centering
    \begin{tabular}{l|ll}
        databases & matches & same composition  \\ \hline
        CSD vs COD & 276,494 & 276,376 \\
        CSD vs ICSD & 3,272 & 3,270 \\
        COD vs ICSD & 35,162 & 32,023 \\
        COD vs MP & 14 & 4 \\
        ICSD vs MP & 71 & 32 \\
    \end{tabular}
    \caption{Number of exact matches ($\EMD$ within $10^{-10}\angstrom$) between the four major databases.}
    \label{tab:cross_db_matches}
\end{table}

\begin{table}[h!]
\caption{These times for all comparisons by COMPACK \cite{chisholm2005compack} are extrapolated 
on the same machine, which completed Table~3 
of near-duplicates across all the major databases within 2 hours.}
\label{tab:COMPACK_times}
\begin{center}
\begin{tabular}{lcrrrr}
database & periodic crystals & unordered pairs & COMPACK time, sec & 
years \\
\hline
CSD & 907,246 & 411,547,198,635 & $4.81\times 10^{10}$ & 
1526 \\
COD & 357,510 & 63,906,521,295 & $7.48\times 10^{9}$ & 
237 \\
ICSD & 170,206 & 14,484,956,115 & $1.69\times 10^{9}$ & 
53 \\
MP & 153,235 & 11,740,405,995 & $1.37\times 10^{9}$ & 
43 \\
GNoME & 384,938 & 74,088,439,453 & $8.67\times 10^{9}$ & 274
\end{tabular}
\end{center}
\end{table}


\begin{figure}
\centering
\includegraphics[width=\textwidth]{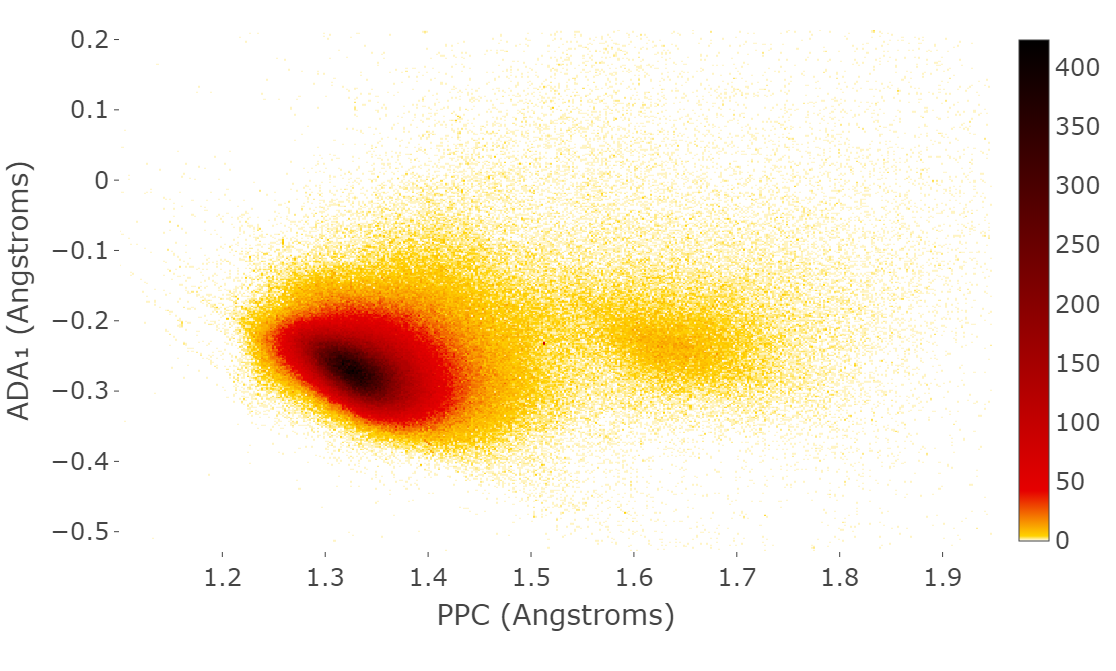}
\includegraphics[width=\textwidth]{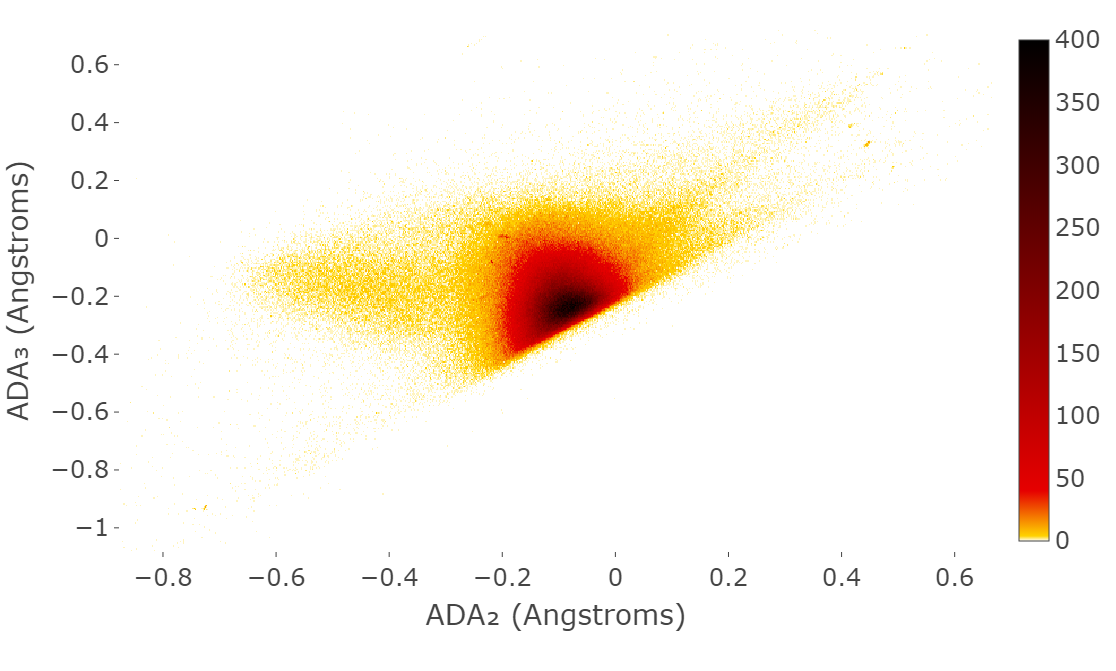}
\caption{The projections of the CSD in the invariants $\PPC,\ADA_1,\ADA_2,\ADA_3$.}
\label{fig:CSD} 
\end{figure}

\begin{figure}
\centering
\includegraphics[width=\textwidth]{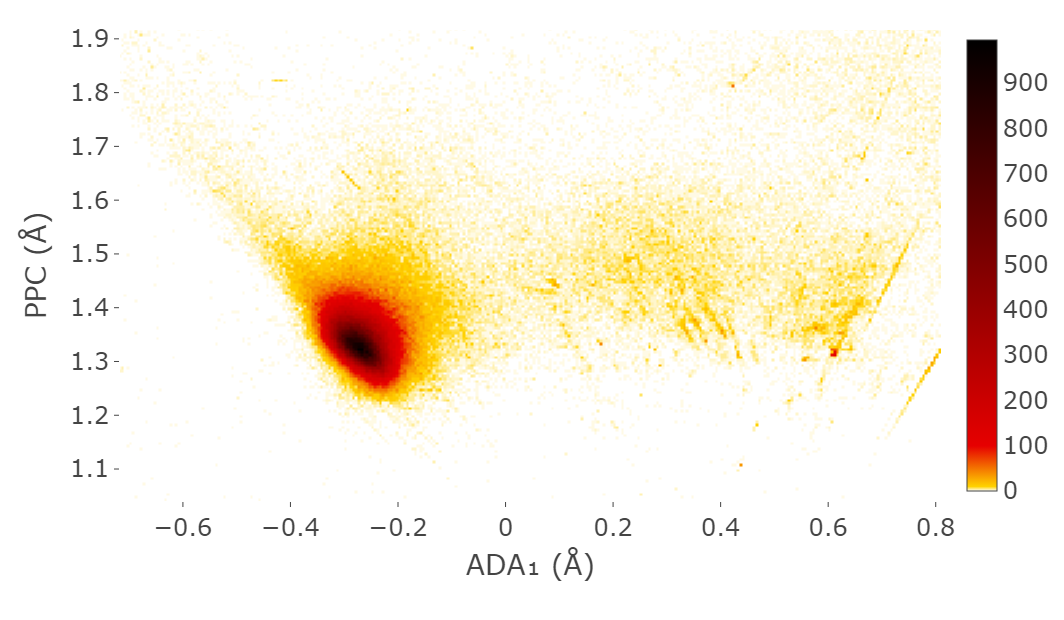}
\includegraphics[width=\textwidth]{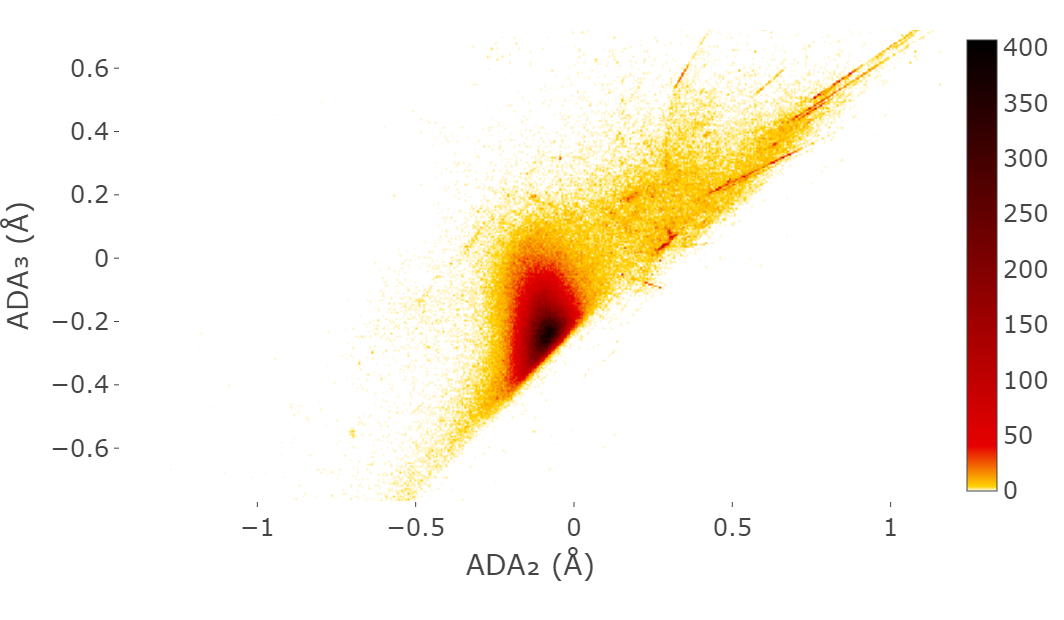}
\caption{The projections of the COD in the invariants $\PPC,\ADA_1,\ADA_2,\ADA_3$.}
\label{fig:COD} 
\end{figure}

\begin{figure}
\centering
\includegraphics[width=\textwidth]{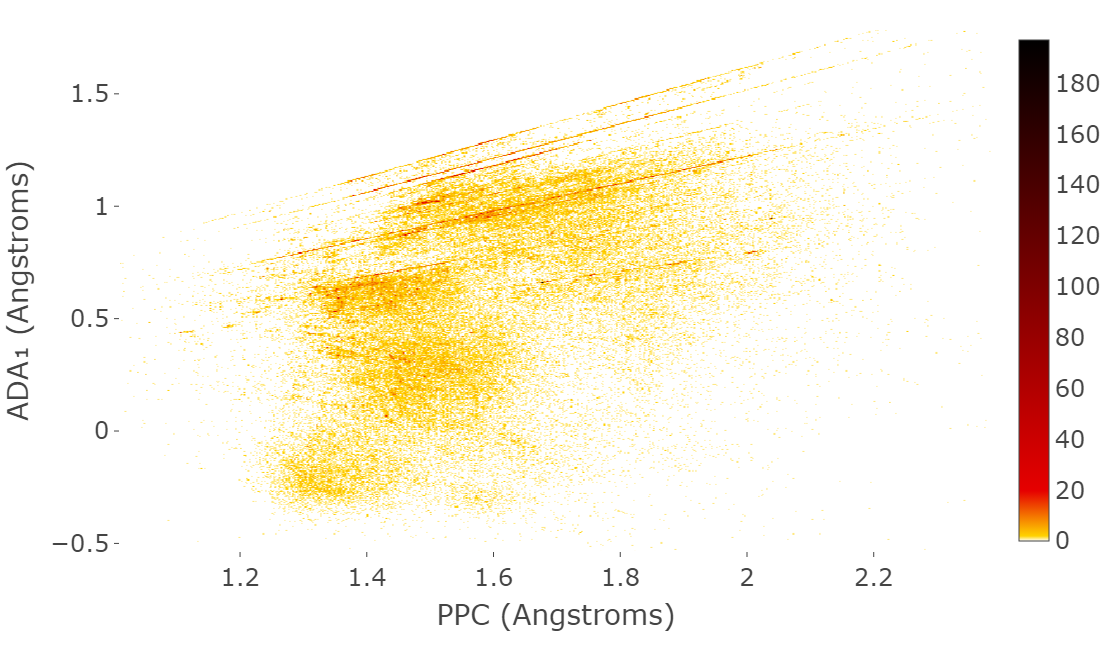}
\includegraphics[width=\textwidth]{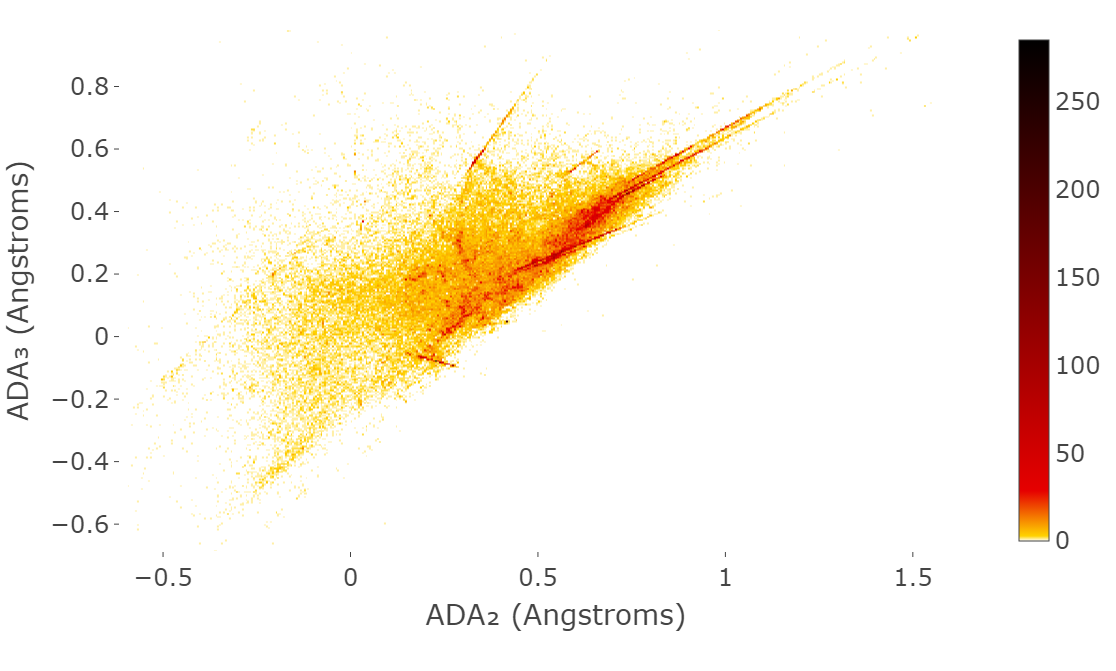}
\caption{The projections of the ICSD in the invariants
$\PPC,\ADA_1,\ADA_2,\ADA_3$.}
\label{fig:ICSD} 
\end{figure}

\begin{figure}
\centering
\includegraphics[width=\textwidth]{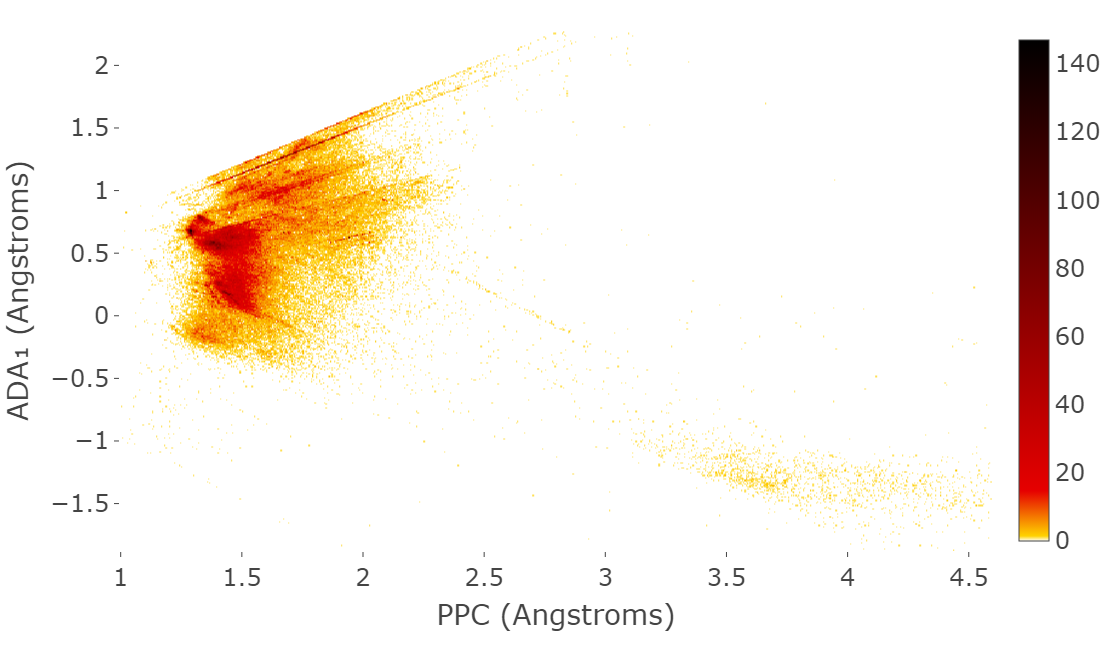}
\includegraphics[width=\textwidth]{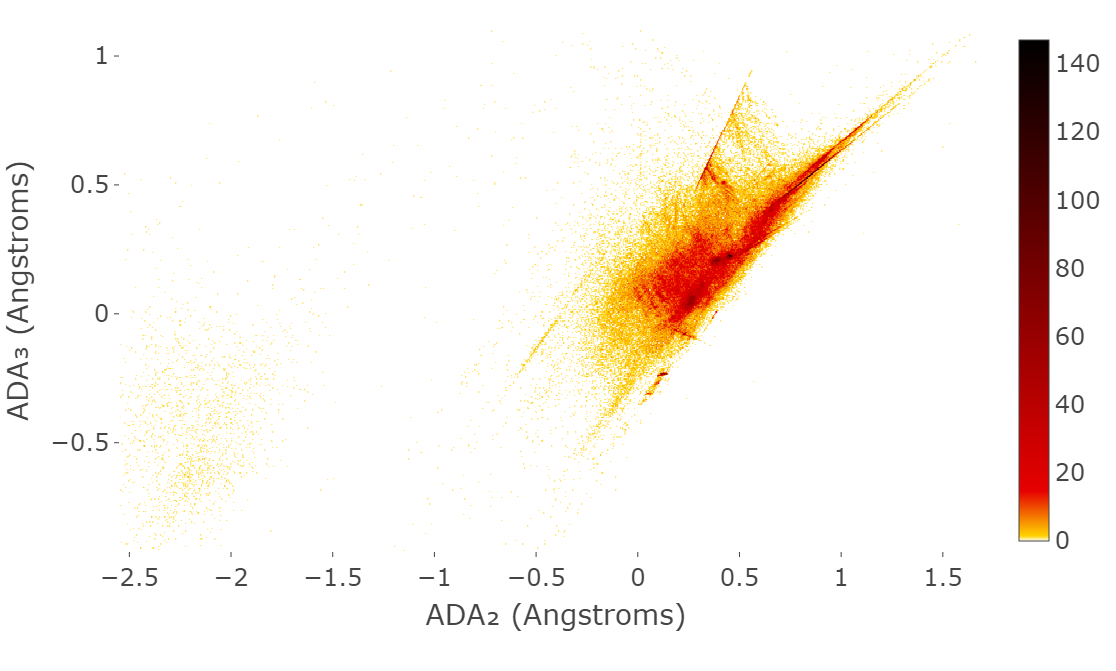}
\caption{The projections of the MP in the invariants $\PPC,\ADA_1,\ADA_2,\ADA_3$.}
\label{fig:MP} 
\end{figure}
\smallskip

Table~\ref{tab:prob_comparison} compares the proven properties of past and new descriptors.

\begin{table}
\centering
\begin{tabular}{lccccc} 
	                 Descriptor                  & Invariant  & Continuity &  Complete   & Reconstruction & Time \\ 
	                 \hline
	             primitive cell                & $\times$ & $\times$ & $\times$  &   $\times$   & \checkmark \\
	                reduced cell                 & \checkmark & $\times$ & $\times$  &   $\times$   & \checkmark \\
	                space group                  & \checkmark & $\times$ & $\times$  &   $\times$   & \checkmark \\
	   PDF~\cite{terban2021structural}    & \checkmark & \checkmark & $\times$  &   $\times$   & \checkmark* \\
	      MACE~\cite{batatia2022mace}                       & \checkmark & $\times$ & \checkmark*  &   $\times$   & \checkmark* \\
	      \hline
	densities~\cite{edelsbrunner2021density} & \checkmark & \checkmark & \checkmark* &   $\times$   & \checkmark* \\
	       AMD \cite{widdowson2022average}      & \checkmark & \checkmark & $\times$  &   $\times$   & \checkmark \\
	       PDD \cite{widdowson2022resolving}     & \checkmark & \checkmark & \checkmark* &  \checkmark*   & \checkmark \\
	       	      isosets~\cite{anosova2021isometry,anosova2026recognition}       & \checkmark & \checkmark & \checkmark  &   \checkmark   & \checkmark* 
\end{tabular}
\caption{Comparison of crystal descriptors in the context of Problem~1.6. 
\checkmark* in the `Computable' column indicates that only an approximate algorithm exists for distances, and \checkmark* in the `Complete' and `Reconstruction' columns means that the condition holds in \emph{general position}.
For example, all invariants based on local atomic environments, such as MACE \cite{batatia2022mace}, discontinuously change under almost any perturbation that arbitrarily scales up a primitive cell as in Fig.~2~(left), except the complete isosets \cite{anosova2021isometry} whose continuous metric was designed to be Lipschitz continuous  \cite{anosova2026recognition}.}
\label{tab:prob_comparison}
\end{table}

\section{Examples and instructions for the PDD code and data}
\label{sec:PDD_ext_code}

This appendix explains the code at https://pypi.org/project/average-minimum-distance.

\subsection{Pseudocode for computing Pointwise Distance Distributions}

The algorithm accepts any periodic point set $S\subset\R^n$ in the form of a unit cell $U$ and a motif $M\subset S$. The cell is given as a square $n\times n$ matrix with basis vectors in the columns, and the motif points in Cartesian form lying inside the unit cell. For dimension 3, the typical Crystallographic Information File (CIF) with six unit cell parameters and motif points in terms of the cell basis is easily converted to this format. Otherwise, the unit cell and motif points can be given directly, in any dimension.
Specifically, the PDD function's interface is as follows:
\smallskip

\noindent Input:
\begin{itemize}
    \item \verb!motif!: array shape $(m,n)$. Coordinates of motif points in Cartesian form.
    \item \verb!cell!: array shape $(n,n)$. Represents the unit cell in Cartesian form.
    \item \verb!k!: \verb!int! $> 0$. Number of columns to return in PDD$(S;k)$.
\end{itemize}

\noindent Output:
\begin{itemize}
    \item \verb|pdd|: array with $k+1$ columns.
\end{itemize}

\smallskip

Before giving the pseudocode, we outline the key objects and functions in use:
\smallskip

\begin{itemize}
    \item A generator \verb!g!, which creates points from the set $S$ to find distances to,
    \item KDTrees (canonically $k$ is the dimension here, in our case it's denoted $n$), data structures designed for fast nearest-neighbor lookup in $\R^n$.
\end{itemize}
\smallskip

Once \verb!g! is constructed, \verb!next(g)! is called to get new points from the infinite set $S$. The first call returns all points in the given unit cell (i.e. the motif), and successive calls returns points from unit cells further from the origin in a spherical fashion.
\smallskip

A KDTree is constructed with a point set $T$, then queried with another $Q$, returning a matrix with distances from all points in $Q$ to their nearest neighbors (up to some given number, $k$ below) in $T$, as well as the indices of these neighbors in $T$.
\smallskip

The functions \verb|collapse_equal_rows| and \verb|lexsort_rows|, which perform the collapsing and lexicographical sorting steps of computing PDD, respectively, are assumed to be implemented elsewhere.
The following pseudocode finds $\PDD(S;k)$ for a periodic set $S$ described by \verb|motif| and \verb|cell|:
\begin{verbatim}
def PDD(motif, cell, k):
    cloud = [] # contains points from S
    g = point_generator(motif, cell)  
    
    # at least k points will be needed 
    while len(cloud) < k:
        points = next(g)
        cloud.extend(points)        
    # first distance query
    tree = KDTree(cloud)
    D_, inds = tree.query(motif, k)
    D = zeros_like(D_)
        
    # repeat until distances don't change, 
    # then all nearest neighbors are found 
    while not D == D_:
        D = D_
        cloud.extend(next(g))
        tree = KDTree(cloud)
        D_, inds = tree.query(motif, k)
    pdd = collapse_equal_rows(D_)
    pdd = lexsort_rows(pdd)  
    return pdd
\end{verbatim}

\subsection{Instructions for the attached PDD code and specific examples}

A Python script implementing Pointwise Distance Distributions along with examples can be found in the zip archive included in this submission. Python 3.7 or greater is required. The dependency packages are NumPy ($<1.22$), SciPy ($\geq1.6.1$), numba ($\geq0.55.0$) and ase ($\geq3.22.0$); if you do not wish to affect any currently installed versions on your machine, create and activate a virtual environment before the following.
\smallskip

Unzip the archive and in a terminal navigate to the unzipped folder. Install the requirements by running \verb|pip install -r requirements.txt|. 
Run \verb|python| followed by the example script of choice, and then any arguments (outlined below), e.g.
\smallskip

\begin{verbatim}
    $ python kite_trapezium_example.py    
    trapezium: [(0, 0), (1, 1), (3, 1), (4, 0)]
    PDD:
    [[0.5        1.41421356 2.         3.16227766]
     [0.5        1.41421356 3.16227766 4.        ]]  
    kite: [(0, 0), (1, 1), (1, -1), (4, 0)]
    PDD:
    [[0.25       1.41421356 1.41421356 4.        ]
     [0.5        1.41421356 2.         3.16227766]
     [0.25       3.16227766 3.16227766 4.        ]] 
    EMD between trapezium and kite: 0.874032
\end{verbatim}
\smallskip

Here is the list of included example scripts and their parameters:
\smallskip

\begin{itemize}

    \item \verb|kite_trapezium_example.py| prints the PDDs of the 4-point sets $K$ (kite) and $T$ (trapezium) in Fig.~\ref{fig:non-isometric_pairs}~(left), along with their EMD.

\begin{figure}
\centering
\includegraphics[height=14.5mm]{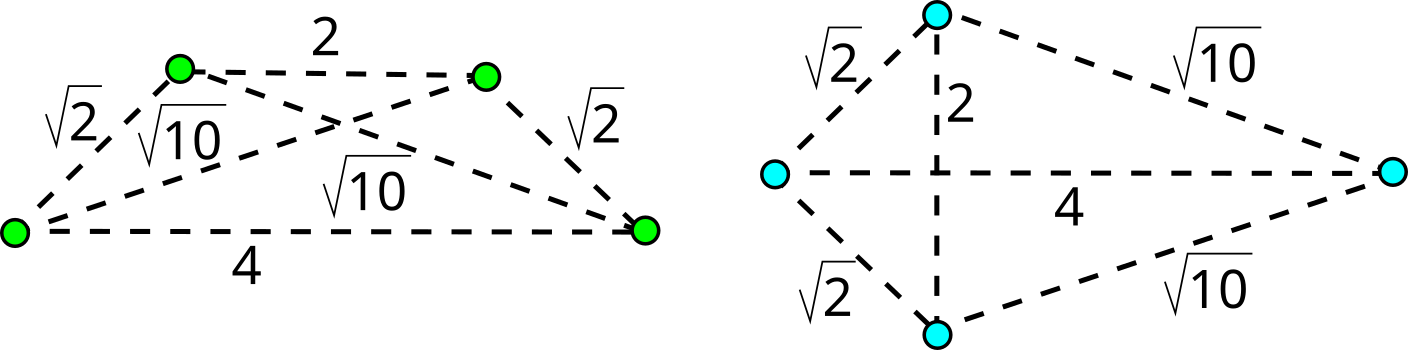}
\includegraphics[height=14.5mm]{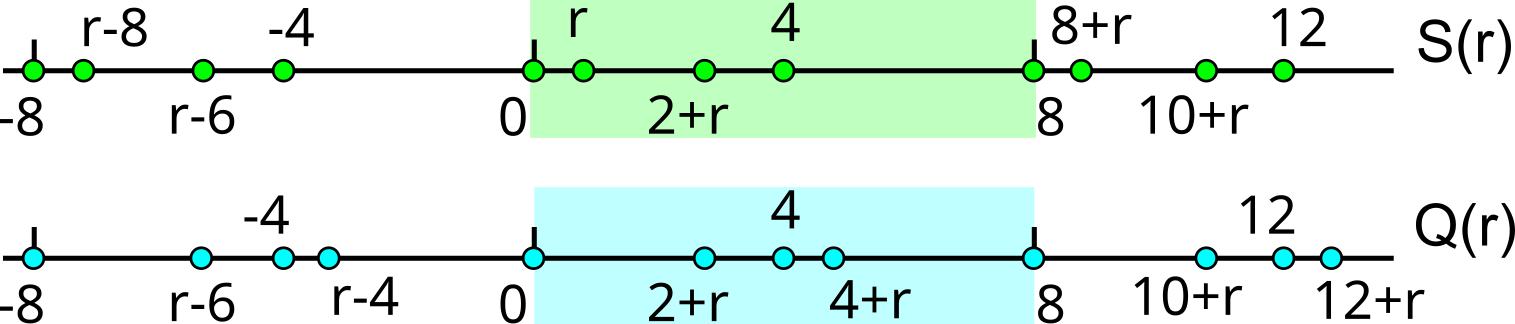}
\caption{\textbf{Left}: the 4-point sets $K=\{(\pm 2,0),\; (\pm 1,1)\}$ and $T=\{(\pm 2,0),(-1,\pm 1)\}$ have the same pairwise distances $\sqrt{2},\sqrt{2},2,\sqrt{10},\sqrt{10},4$.
\textbf{Right}: the sequences $S(r)=\{0,r,2+r,4\}+8\Z$ and $Q(r)=\{0,2+r,4,4+r\}+8\Z$ for $0<r\leq 1$ have the same Patterson
function \cite[p.~197, Fig.~2]{patterson1944ambiguities}.}
\label{fig:non-isometric_pairs}
\end{figure}

    \item \verb|1D_sets_example.py| shows that the 1D periodic sets in  Fig.~\ref{fig:non-isometric_pairs}~(right) are distinguished by their PDDs for any $0< r \leq 1$. This script requires $r$ to be passed after the file name, e.g. `\verb|python 1D_sets_example.py 0.5|'.
    
    \item \verb|T2_14_15_example.py| compares the crystals shown in Fig.~\ref{fig:T2_14_15}, whose original CIFs are included. This optionally accepts the number $k$ of columns in the computed PDD, e.g.  `\verb|python T2_14_15_example.py --k 50|' compares by PDD with $k=50$. If not included, $k=100$ is used as the default.
 
 \newcommand{\CSPh}{42mm}
\begin{figure}
\centering
\includegraphics[height=\CSPh]{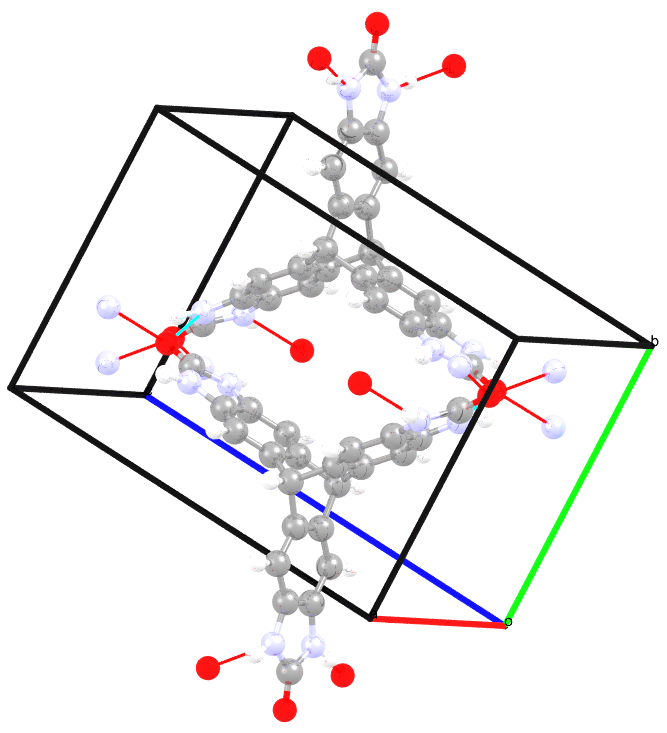}
\includegraphics[height=\CSPh]{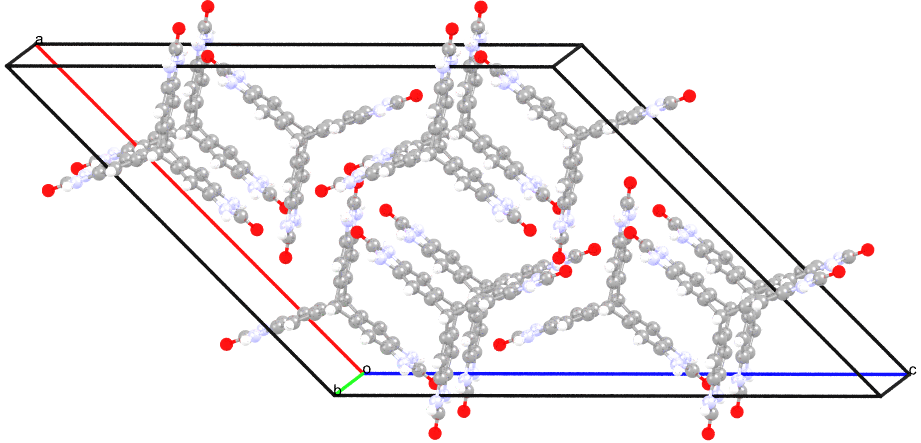}
\caption{Crystals 14, 15 from the database of 5679 simulated crystals reported in  \cite{pulido2017functional} consist of identical T2 molecules and have very different Crystallographic Information Files (with different motifs in unit cells of distinct shapes) but are nearly identical under isometry.}
\label{fig:T2_14_15}
\end{figure}
       
    \item \verb|CSD_duplicates_example.py| computes and compares the PDDs of isometric crystals from the CSD discussed in section~\ref{sec:exp_details}, giving distances of exactly zero. This optionally accepts the parameter $k$ controlling the number of columns in the computed PDD, in the same way as \verb|T2_14_15_example.py|.
    
\end{itemize}
\smallskip

\begin{figure}
\centering
\includegraphics[height=45mm]{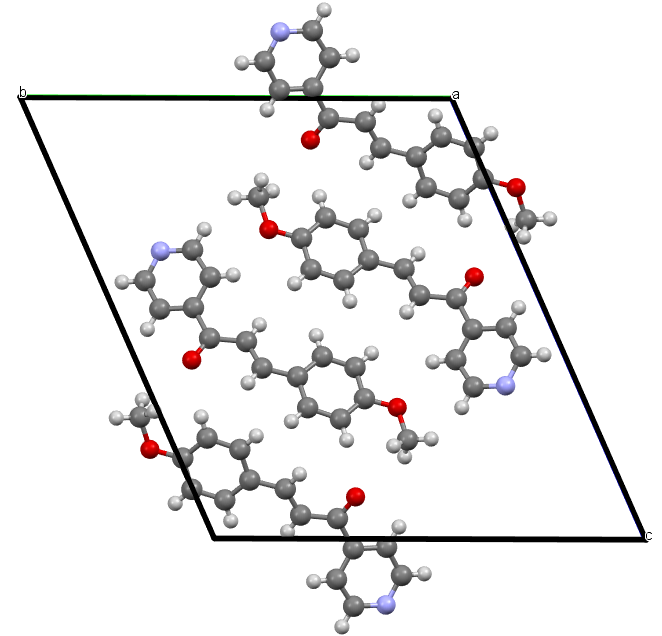}
\includegraphics[height=45mm]{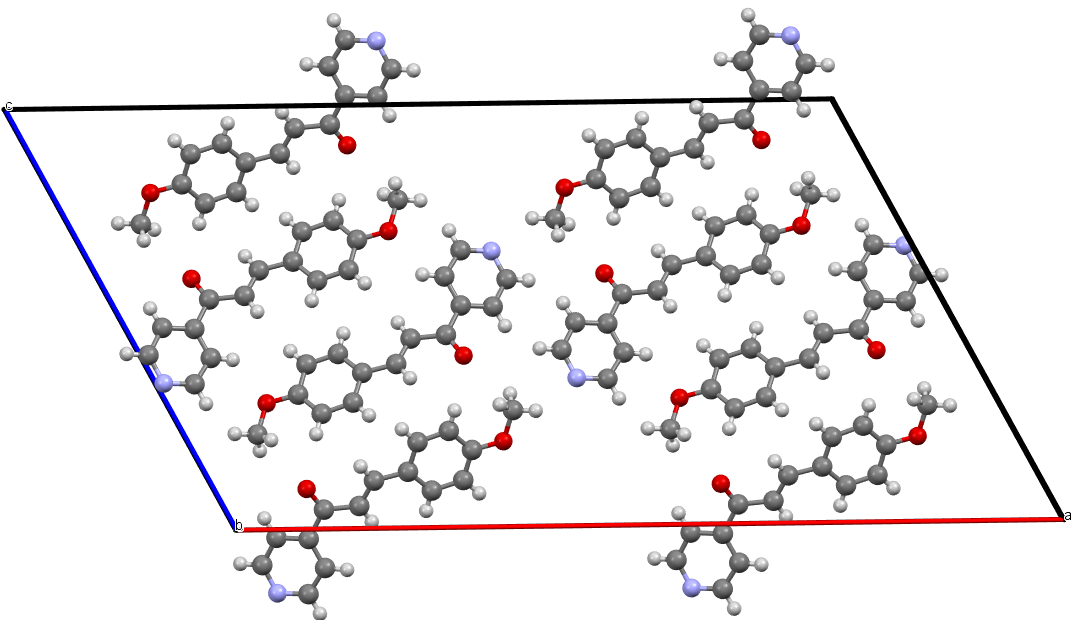}
\caption{In the CSD, near-duplicates PUBTEM (left) and PUBTEM01 (right) have a very small $\EMD= 0.00038\angstrom$ on invariants $\PDA(S;100)$, though their unit cells are rather different.}
\label{fig:CSD_near-duplicates_diff_cells}
\end{figure}

\begin{figure}
\centering
\includegraphics[width=\textwidth]{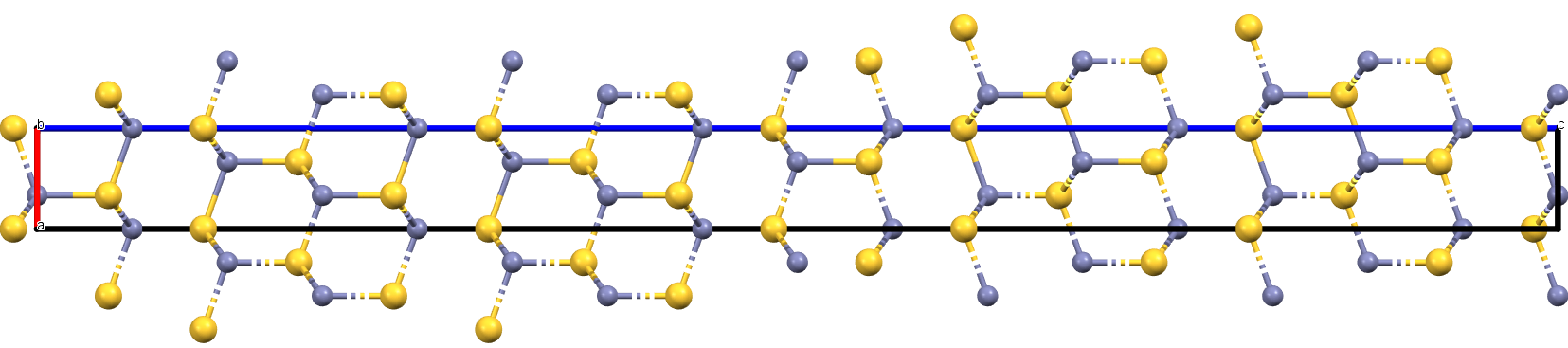}
\includegraphics[width=\textwidth]{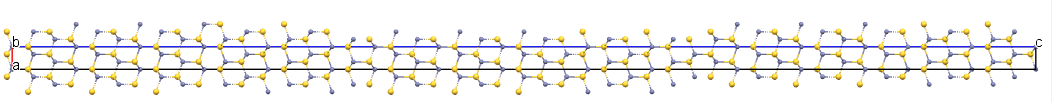}
\caption{In the COD, near-duplicates 2310812 (top) and 2310813 (bottom) have a very small $\EMD= 0.0008\angstrom$ on invariants $\PDA(S;100)$, though their unit  cells differ by a factor of about 3.}
\label{fig:COD_near-duplicates_diff_cells}
\end{figure}

\begin{figure}
\centering
\includegraphics[width=\textwidth]{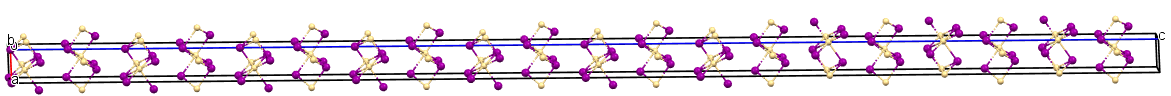}
\includegraphics[width=\textwidth]{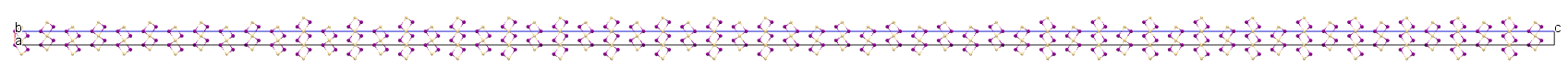}
\caption{In the ICSD, near-duplicates 42291 (top) and 42302 (bottom) have a very small $\EMD= 0.0024\angstrom$ on invariants $\PDA(S;100)$, though their unit  cells differ by a factor of about 3.}
\label{fig:ICSD_near-duplicates_diff_cells}
\end{figure}

\begin{figure}
\centering
\includegraphics[height=15mm]{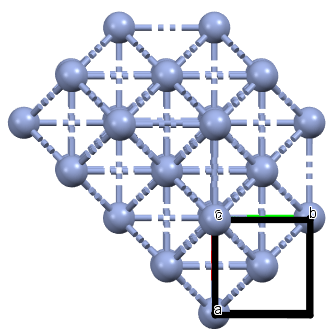}
\includegraphics[height=15mm]{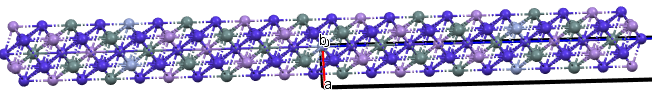}
\caption{In the MP, near-duplicate entries mp-90 (left) and mp-1221808 (right) have a very small $\EMD= 0.0087\angstrom$ on invariants $\PDA(S;100)$, though their unit cells substantially differ.}
\label{fig:MP_near-duplicates_diff_cells}
\end{figure}

\begin{figure}
\centering
\includegraphics[height=45mm]{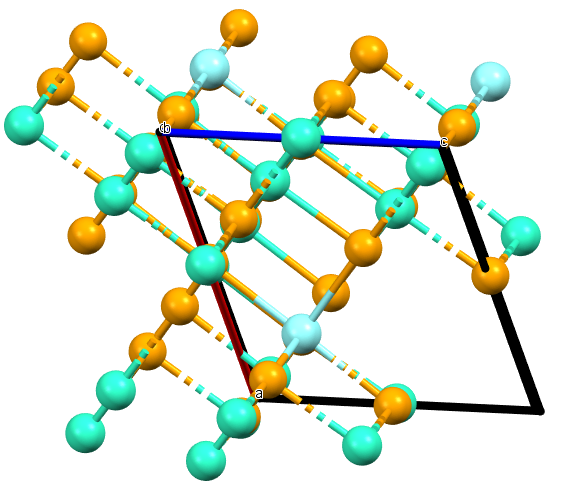}
\includegraphics[height=45mm]{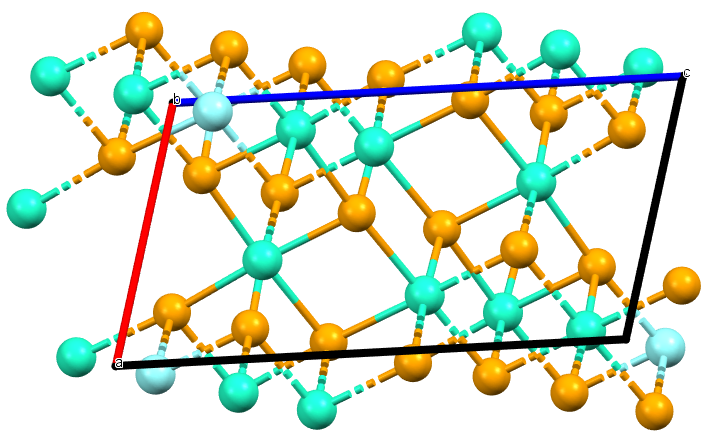}
\caption{In the GNoME, near-duplicates 4cb3b6ed9f (left) and 776c1b7570 (right) in the GNoME have $\EMD= 0.0079\angstrom$ on invariants $\PDA(S;100)$, though their unit cells are very different.}
\label{fig:GNoME_near-duplicates_diff_cells}
\end{figure}

If you wish to run the code on your own sets or CIF files, you can use the functions exposed in the main script \verb|pdd.py|. Use \verb|pdd.read_cif()| to parse a cif and return a crystal, or define one manually as a tuple \verb|(motif, cell)| with NumPy arrays. Pass this as the first argument to \verb|pdd.pdd()| with an integer \verb|k| as the second to compute the PDD. Pass two PDDs to \verb|pdd.emd()| to calculate the Earth mover's distance between them. For finite sets, the function \verb|pdd.pdd_finite()| accepts just one argument, an array containing the points, and returns the PDD.
Figures
\ref{fig:CSD_near-duplicates_diff_cells},
\ref{fig:COD_near-duplicates_diff_cells},
\ref{fig:ICSD_near-duplicates_diff_cells},
\ref{fig:MP_near-duplicates_diff_cells},
\ref{fig:GNoME_near-duplicates_diff_cells},
 show near-duplicates with very different cells, which were counted in Table~4. 

\section{Detailed proofs of auxiliary lemmas and Theorem 4.2} 
\label{sec:PDD_ext_proofs}

This appendix proves Lemmas 3.4-3.5, which were used in Theorem 3.6, and Theorem 4.2.

\begin{proof}[Proof of Lemma~3.4] 
Intersect the three regions ${U^-}(p;r)\subset C(p;r)\subset {U^+}(p;r)$ with $S$ in $\R^n$ and count all points:
$|S\cap{U^-}(p;r)|\leq |S\cap C(p;r)|\leq |S\cap{U^+}(p;r)|$.
\smallskip

The union ${U^-}(p;r)$ consists of $m^{-}(p;r)=\dfrac{\vol[U^-(p;r)\cap R^l]}{\vol[U]}$ shifted cells, which all have the same volume $\vol[U\cap R^l]$.
Since $|S\cap U|=m$, we get $|S\cap U^-(p;r)|=\dfrac{\vol[U^-(p;r)\cap R^l]}{\vol[U]}m$.
Similarly, we count all points of $S$ in the upper union as follows:
$|S\cap{U^+}(p;r)|=\dfrac{\vol[{U^+}(p;r)\cap R^l]}{\vol[U]}m$.
The bounds for $|S\cap C(p;r)|$ become
$$\dfrac{\vol[{U^-}(p;r)\cap R^l]}{\vol[U]}m
\leq|S\cap C(p;r)|\leq
\dfrac{\vol[{U^+}(p;r)\cap R^l]}{\vol[U]}m,$$
which proves the internal inequalities 
$m^{-}(p;r) m\leq |S\cap C(p;r)|\leq m^{+}(p;r)m$.
Then
$$\vol[{U^-}(p;r)\cap R^l]\leq\dfrac{\vol[U\cap R^l]}{m}|S\cap C(p;r)|\leq\vol[{U^+}(p;r)\cap R^l].$$
For the width $w$ of the unit cell $U$, the smaller cylinder $C(p;r-w)$ is completely contained within the lower union ${U^-}(p;r)$.
Indeed, if $|\vec q-\vec p|\leq r-w$, then $q\in U+\vec v$ for some $\vec v\in\La$.
Then $(U+\vec v)$ is covered by the cylinder $C(q;w)$, hence by $C(p;r)$ due to the triangle inequality.
The inclusion $C(p;r-w)\subset{U^-}(p;r)$ implies the lower bound for the volumes: 
$(r-w)^l V_l=\vol[C(p;r-w)\cap R^l]\leq\vol[{U^-}(p;r)\cap R^l]$, where $V_l$ is the unit ball volume in $\R^l$.
Then $\dfrac{(r-w)^l V_l}{\vol[U\cap R^l]}\leq
\dfrac{\vol[{U^-}(p;r)\cap R^l]}{\vol[U\cap R^l]}=m^-(p;r)$,
which implies the first required inequality in the lemma:
$$\left(\dfrac{r-w}{\PPC(S)}\right)^l
=\dfrac{(r-w)^l m V_l}{\vol[U\cap R^l]}
\leq
\dfrac{\vol[{U^-}(p;r)\cap R^l]}{\vol[U\cap R^l]}m=m^{-}(p;r) m.$$
The last required inequality is proved similarly by using $U^+(p;r)\subset C(p;r+w)$.
\end{proof}
\smallskip

\begin{proof}[Proof of Lemma~3.5] 
Let $q\in S$ be a $k$-th neighbor of $p$ in $S$.
There can be several points $q\in S$ at the distance $|q-p|=d_k(S;p)$ but the argument below works for any $q$.
The closed cylinder $C(p;r)$ with $r=d_k(S;p)$ contains the $k$-th neighbor $q$ of $p$ and hence has more than $k$ points (including $p$) from $S$.
The upper bound of Lemma~3.4 
for $r=d_k(S;p)$ implies that $k<|S\cap C(p;r)|\leq\dfrac{(r+w)^l}{(\PPC(S))^l}$.
Taking the $l$-th roots gives $\sqrt[l]{k}<\dfrac{r+w}{\PPC(S)}$, so
$r=d_k(S;p)>\PPC(S)\sqrt[l]{k}-w$. 
\smallskip

For any radius $r$ such that $\sqrt{r^2+h^2}<d_k(S;p)$, the closed cylinder $C(p;r)$ contains only points at a maximum distance $\sqrt{r^2+h^2}$ from $p$.
Then $C(p;r)$ does not include the $k$-th neighbor $q$ of $p$ and hence contains at most $k$ points (including $p$) from $S$.
The lower bound of Lemma~3.4 
for $r<\sqrt{(d_k(S;p))^2-h^2}$ implies that $\dfrac{(r-w)^l}{(\PPC(S))^l}\leq |S\cap C(p;r)|\leq k$.
Since the inequality $\dfrac{(r-w)^l}{(\PPC(S))^l}\leq k$ holds for the constant upper bound $k$ and any radius $r<\sqrt{(d_k(S;p))^2-h^2}$, the same inequality holds for the radius $r=\sqrt{(d_k(S;p))^2-h^2}$.
Then
$\dfrac{r-w}{\PPC(S)}\leq\sqrt[l]{k}$, 
$$r=\sqrt{(d_k(S;p))^2-h^2}\leq \PPC(S)\sqrt[l]{k}+w, \quad
d_k(S;p)\leq \sqrt{(\PPC(S)\sqrt[l]{k}+w)^2+h^2}.$$ 
\end{proof}
\smallskip

\begin{exa}[stronger asymptotic $\ADA_k(S)\to 0$ as $k\to+\infty$ for $\Z^n$]
\label{exa:cubic_lattice_asymptotic}
The survey \cite{ivic2004lattice} describes progress on the generalized Gauss circle problem expressing the number of points from the cubic lattice $\Z^n$ within a ball of a radius $r$ as $k=V_n r^n-O(r^{\al_n+\ep})$ for any $\ep>0$, where 
$\al_n<n-1$ for $n\geq 2$.
The cubic lattice has $\PPC(\Z^n)=1/\sqrt[n]{V_n}$.
Let $d_k$ denote the distance from the origin $0$ to its $k$-th neighbor in $\Z^n$.
Then $k=V_n d_k^n-O(d_k^{\al_n+\ep})$, so
 $d_k = \sqrt[n]{\dfrac{k+O(d_k^{\al_n+\ep})}{V_n}}=\PPC(\Z^n)\sqrt[n]{k+O(d_k^{\al_n+\ep})}$.
Then
$$\dfrac{\ADA_k(\Z^n)}{\PPC(\Z^n)}=\dfrac{d_k}{\PPC(\Z^n)}-\sqrt[n]{k}=\sqrt[n]{k+O(d_k^{\al_n+\ep})}-\sqrt[n]{k}
=\dfrac{O(d_k^{\al_n+\ep})}{P_n(\sqrt[n]{k+O(d_k^{\al_n+\ep})},\sqrt[n]{k})},$$ where $P_n$ is a homogeneous polynomial of degree $n-1$, e.g. $P_2(x,y)=x+y$, $P_3(x,y)=x^2+xy+y^2$. 
Since the numerator has the power $\al_n<n-1$ of $d_k=O(\sqrt[n]{k})$ for $n\geq 2$, the final expression and hence $\ADA_k(\Z^n)$ have limit $0$ as $k\to+\infty$.
\end{exa}
\smallskip

Theorem~4.1 
will be proved similar to \cite[Theorem~13]{widdowson2022average} 
by Lemmas~\ref{lem:common_lattice}, \ref{lem:perturbed_distances}, \ref{lem:perturbed_vectors}.
Partial cases of Lemmas~\ref{lem:common_lattice} and \ref{lem:perturbed_distances} appeared for $l=n$ in \cite[Lemma~2]{edelsbrunner2021density} and for $\R^n$ in \cite[Lemma 8]{widdowson2022average}, respectively.
\smallskip

\begin{lem}[common lattice]
\label{lem:common_lattice}
Let $l$-periodic point sets $S,Q\subset\R^n$ have a bottleneck distance  $d_B(S,Q)<\min\{r(S),r(Q)\}$. 
Then $S,Q$ have a common lattice $\La$ with a unit cell $U$ such that
$S=\La+(U\cap S)$ and $Q=\La+(U\cap Q)$.
\end{lem}
\newcommand{\lemcommonlatticeproof}{
\begin{proof}[Proof of Lemma~\ref{lem:common_lattice}]
Choose the origin $0\in\R^n$ at a point of $S$.
Applying translations, we can assume that primitive unit cells $U(S),U(Q)$ of the given $l$-periodic sets $S,Q$ have a vertex at the origin $0$.
Then $S=\La(S)+(U(S)\cap S)$ and $Q=\La(Q)+(U(Q)\cap Q)$, where 
$\La(S),\La(Q)$ are $l$-dimensional lattices of $S,Q$, respectively.
We are given that every point of $Q$ is $d_B(S,Q)$-close to a point of $S$, where the bottleneck distance $d_B(S,Q)$ is strictly less than the packing radius $r(Q)$.  
\smallskip

Assume by contradiction that $S,Q$ have no common lattice.
Then there is a point $p\in\La(S)\subset S$ whose all integer multiples $k\vec p\in\La(S)$ do not belong to $\La(Q)$ for $k\in\Z-\{0\}$.
Any such multiple $k\vec p\in\La(S)\subset S$ can be translated by a vector of $\La(Q)$ to a point $t(k)$ in the unit cell $U(Q)$ so that $k\vec p\equiv t(k)\pmod{\La(Q)}$.
Since the cell $U(Q)$ contains infinitely many points $t(k)$ for $k\neq 0$,
one can find a pair $t(i)\neq t(j)$ at a distance less than $\de=r(Q)-d_B(S,Q)>0$.
For any $m\in\Z$, the following points are equivalent modulo (translations along the vectors of) the lattice $\La(Q)$.
$$t(i+m(j-i))\equiv (i+m(j-i))\vec p= 
 i\vec p + m(j\vec p-i\vec p)\equiv 
 t(i) + m(t(j)-t(i)).$$
These points for $m\in\Z$ lie in a straight line with gaps $|t(j)-t(i)|<\de$.
The open balls with the packing radius $r(Q)$ and centers at all points of $Q$ do not overlap.
Hence all closed balls with the radius $d_B(S,Q)<r(Q)$ and the same centers are at least $2\de$ away from each other.
Due to $|t(j)-t(i)|<\de=r(Q)-d_B(S,Q)$, there is $m\in\Z$ such that $t(i) + m(t(j)-t(i))$ is outside the union $Q+\bar B(0;d_B(S,Q))$ of all these smaller balls.
Then $t(i) + m(t(j)-t(i))$ has a distance more than $d_B(S,Q)$ from any point of $Q$. 
The translations along all vectors of the lattice $\La(Q)$ preserve the union of balls $Q+\bar B(0;d_B(S,Q))$.
Then the point $(i+m(j-i))\vec p\in\La(S)\subset S$, which is equivalent to $t(i) + m(t(j)-t(i))$ modulo $\La(Q)$, has a distance more than $d_B(S,Q)$ from any point of $Q$.
This conclusion contradicts the definition of 
$d_B(S,Q)$. 
\end{proof}
}\lemcommonlatticeproof

\begin{lem}[perturbed distances]
\label{lem:perturbed_distances}
For some $\ep>0$, let $g:S\to Q$ be a bijection between any discrete sets in a space $X$ with a metric $d_X$ such that $d_X(g(p),p)\leq\ep$ for all $p\in S$.
Then, for any $i\geq 1$, let $p_i\in S$, $\ti p_i\in Q$ be the $i$-th nearest neighbors of $p\in S$, $\ti p=g(p)\in Q$, respectively.
Then the distances from the points $p,\ti p$ to their $i$-th neighbors $p_i,\ti p_i$ in $X$ are $2\ep$-close to each other, i.e. $|d_X(p,p_i)-d_X(\ti p,\ti p_i)|\leq 2\ep$. 
\end{lem}
\newcommand{\lemperturbeddistancesproof}{
\begin{proof}[Proof of Lemma~\ref{lem:perturbed_distances}]
Shifting the point $g(p)$ back to $p$, assume that $p=g(p)$ is fixed and all other points change their positions by at most $2\ep$.
Assume by contradiction that the distance from $p$ to its new $i$-th neighbor $t_i$ is less than $d_X(p,p_i)-2\ep$.
Then all first new $i$ neighbors $\ti p_1,\dots,\ti p_i\in Q$ of $p$ belong to the open ball with the center $p$ and radius $d_X(p,p_i)-2\ep$. 
Since the bijection $g$ shifted every $\ti p_1,\dots,\ti p_i$ by at most $2\ep$, their preimages $g^{-1}(\ti p_1),\dots,g^{-1}(\ti p_i)$ belong to the open ball with the center $p$ and the radius $d_X(p,p_i)$.
Then the $i$-th neighbor of $p$ within $S$ is among these $i$ preimages, i.e. the distance from $p$ to its $i$-th nearest neighbor should be strictly less than the assumed value 
$d_X(p,p_i)$.
We similarly get a contradiction by assuming that the distance from $p$ to its new $i$-th neighbor $\ti p_i$ is more than $d_X(p,p_i)+2\ep$.
\end{proof}
}\lemperturbeddistancesproof

\begin{lem}[perturbed distance vectors]
\label{lem:perturbed_vectors}
For $\ep>0$, let $g:S\to Q$ be a bijection between any discrete sets in a space $X$ with a metric $d_X$ so that $d_X(g(p),p)\leq\ep$ for all $p\in S$.
Then 
$g$ changes the vector $\vec R(S,p)=(d_X(p,p_1),\dots,d_X(p,p_k))$ of the first $k$ minimum distances from any point $p\in S$ to its $k$ nearest neighbors $p_1,\dots,p_k\in S$ by at most $2\ep\sqrt[q]{k}$ in the distance $L_{q}$.
So if 
$\vec R(Q,\ti p)=(d_X(\ti p,\ti p_1),\dots,d_X(\ti p,\ti p_k))$ is the vector of the first $k$ minimum distances from $\ti p=g(p)$ to its $k$ nearest neighbors $\ti p_1,\dots,\ti p_k$ in $Q$, then $L_{q}(\vec R(S,p),\vec R(Q,\ti p))\leq 2\ep\sqrt[q]{k}$. 
\end{lem}
\begin{proof}[Proof of Lemma~\ref{lem:perturbed_vectors}]
By Lemma~\ref{lem:perturbed_distances}, every coordinate of $\vec R(S,p)$ changes by at most $2\ep$.
Hence the distance $L_{q}(\vec R(S,p),\vec R(Q,\ti p))\leq\big(\sum\limits_{i=1}^k(2\ep)^q\big)^{1/q}=2\ep\sqrt[q]{k}$.   
\end{proof}

\begin{proof}[Proof of Theorem~4.2] 
The bottleneck distance  between the given sets $S,Q\subset X$ is $d_B(S,Q)=\inf\limits_{g:S\to Q}\; \sup\limits_{p\in S}d_X(g(p),p)$.
Then for any $\de>0$ there is a bijection $g:S\to Q$ such that $\sup\limits_{p\in S}d_X(g(p),p)\leq d_B(S,Q)+\de$.
If the given sets $S,Q$ are finite, one can set $\de=0$.
Indeed, there are only finitely many bijections $g:S\to Q$, hence the infimum in the definition above is achieved for one of these bijection $g$.
\smallskip
 
(a) For any discrete sets $S,Q\subset X$ be with finite subsets $M,T$ of the same number $m$ of points, respectively, we use the notations of Definition~3.1. 
The given 1-1 perturbation $g:S\to Q$ defines the simplest 1-1 flow from the row of any $p\in M$ in the matrix $D(S,M;k)$ to the row of $g(p)\in T$ in $D(Q,T;k)$ by setting $f_{ii}=\frac{1}{m}$ and $f_{ij}=0$ for $i\neq j$, where $i,j=1,\dots,m$.
All rows of $D(S,M;k)$ that are identical to each other are collapsed to a single row, similarly for $D(Q,T;k)$.
By summing up weights of all collapsed rows, the above flow induces a flow from all distance vectors in $\PDD(S,M;k)$ to all distance vectors in $\PDD(Q,T;k)$. 
\smallskip

Then $\EMD_q(\PDD(S,M;k),\PDD(Q,T;k))\leq\frac{1}{m}\sum\limits_{i=1}^m L_q(\vec R_i(S), \vec R_i(Q))$, because $\EMD_q$ minimizes the cost in Definition~4.2. 
The upper bound $L_q(\vec R_i(S),\vec R_i(Q))\leq 2(\ep+\de)\sqrt[q]{k}$ from Lemma~\ref{lem:perturbed_vectors} implies that 
$$\EMD_q(\PDD(S,M;k),\PDD(Q,T;k))\leq\frac{1}{m}\sum\limits_{i=1}^m 2(\ep+\de)\sqrt[q]{k}=2(\ep+\de)\sqrt[q]{k},$$ which holds for any small $\de>0$.
By taking the limit for $\de\to 0$, we get the required upper bound 
$\EMD_q(\PDD(S,M;k),\PDD(Q,T;k))\leq 2\ep\sqrt[q]{k}$. 
\smallskip
 
(b) In the $l$-periodic case by Lemma~\ref{lem:common_lattice}, the given sets $S,Q$ should have a common $l$-dimensional lattice $\La$.
Any primitive cell $U$ of $\La$ is a common unit cell of $S,Q$, i.e. $S=\La+(S\cap U)$ and $Q=\La+(Q\cap U)$, so $\PPC(S)=\PPC(Q)$.
Then all $L_\infty$ distances between rows in $\PDA(S;k),\PDA(Q;k)$ are the same as between the corresponding rows in $\PDD(S;k),\PDD(Q;k)$, see Definition~3.7. 
Hence $\EMD_q(\PDA(S;k),\PDA(Q;k))=\EMD_q(\PDD(S;k),\PDD(Q;k))\leq 2\ep\sqrt[q]{k}$ by (a).
\smallskip

The remaining inequality follows from the $\PDA$ case.
Indeed, each element of $\PND(S;k)$ in a row $i$ and a column $j=1,\dots,k$ is obtained from the corresponding element of $\PDA(S;k)$ by dividing by $\PPC(S)\sqrt[l]{j}\geq\PPC(S)$.
Then each distance $L_q$ between corresponding rows in $\PND(S;k)$, $\PND(Q;k)$ is at least $\PPC(S)$ times smaller than between the same rows 
in $\PDA(S;k)$, $\PDA(Q;k)$.
Then
\begin{align*}
& \EMD_q(\PND(S;k),\PND(Q;k))\leq 
\dfrac{\EMD_q(\PDA(S;k),\PDA(Q;k))}{\PPC(S)}\leq 
\dfrac{2\ep\sqrt[q]{k}}{\PPC(S)}.
\end{align*}
\end{proof}
\smallskip

\newcommand{\thmlowerboundproof}{
\begin{proof}[Proof of Theorem~4.4] 
Considering $\PDD(S;k)$ as a weighted distribution of rows, $\AMD(S;k)$ is its centroid from \cite[section~3]{cohen1997earth}.
The argument below follows the proof for $q=+\infty$ of \cite[Theorem~1]{cohen1997earth} and similarly works for other invariants in parts (b,c).
In the notations of Definition~4.1, 
we use the inequality $||\vec u||_q + ||\vec v||_q|| \geq ||\vec u+\vec v||_q$ for the $q$-norm $||\vec v||_q=\big(\sum\limits_{i=1} |v_i|^q \big)^{1/q}$ of the Minkowski metric $L_q$ as follows:
\begin{align*}
& \EMD_q(\PDD(S;k),\PDD(Q;k))=
\sum\limits_{i=1}^{m(S)} \sum\limits_{j=1}^{m(Q)} f_{ij} L_q(\vec R_i(S),\vec R_j(Q))  =\\
& 
\sum\limits_{i=1}^{m(S)} \sum\limits_{j=1}^{m(Q)} ||f_{ij} \big(\vec R_i(S) -\vec R_j(Q)\big)||_q \geq
||\sum\limits_{i=1}^{m(S)} \sum\limits_{j=1}^{m(Q)} f_{ij} (\vec R_i(S) -\vec R_j(Q)) ||_q = \\
& || \sum\limits_{i=1}^{m(S)} \big(\sum\limits_{j=1}^{m(Q)}
f_{ij} \vec R_i(S) \big)-
\sum\limits_{j=1}^{m(Q)} \big(\sum\limits_{i=1}^{m(S)} f_{ij} \vec R_j(Q) \big) ||_q = \\
&
|| \sum\limits_{i=1}^{m(S)} w_i(S) \vec R_i(S) -
\sum\limits_{j=1}^{m(Q)} w_j(Q) \vec R_j(Q) ||_q =
L_q(\AMD(S;k), \AMD(Q;k)).
\end{align*}
\end{proof}
}\thmlowerboundproof

Many authors considered criteria or complete invariants of congruence for plane quadrilaterals
\cite{vance1982minimum}, whose vertices are ordered.
A complete and continuous invariant of $m$ ordered points under isometry in $\R^n$ is the $m\times m$ matrix of pairwise distance \cite{schoenberg1935remarks} or the Gram matrix of scalar products \cite{weyl1946classical}.
The extension of this approach to $m$ unordered points leads to the exponential complexity because of $m!$ permutations. 
For $m=4$ unordered points, Theorem~5.3 proves the completeness of $\PDD(C;m-1)$ under isometry in any $\R^n$.
For any $m$, the invariant $\PDD(C;m-1)$ can computed in quadratic time $O(m^2)$.
For $m=4$, $\PDD(C;3)$ contains only 12 numbers (6 pairs of distances between 4 points), while $4!=24$ distance matrices on 4 points contain at least 144 numbers if we take only distances above the diagonals. 
\medskip

If a cloud $C$ of $m$ points has a line or plane of symmetry $L$ in $\R^2$ or $\R^3$, then all points $C\setminus L$ split into pairs of points that are symmetric in $L$ and hence have equal rows in $\PDD(C;m-1)$.
Lemma~\ref{lem:PDD_symmetry} shows that the converse holds for $m=4$.
\smallskip
 
\begin{figure}[h!]
\centering
\includegraphics[width=\textwidth]{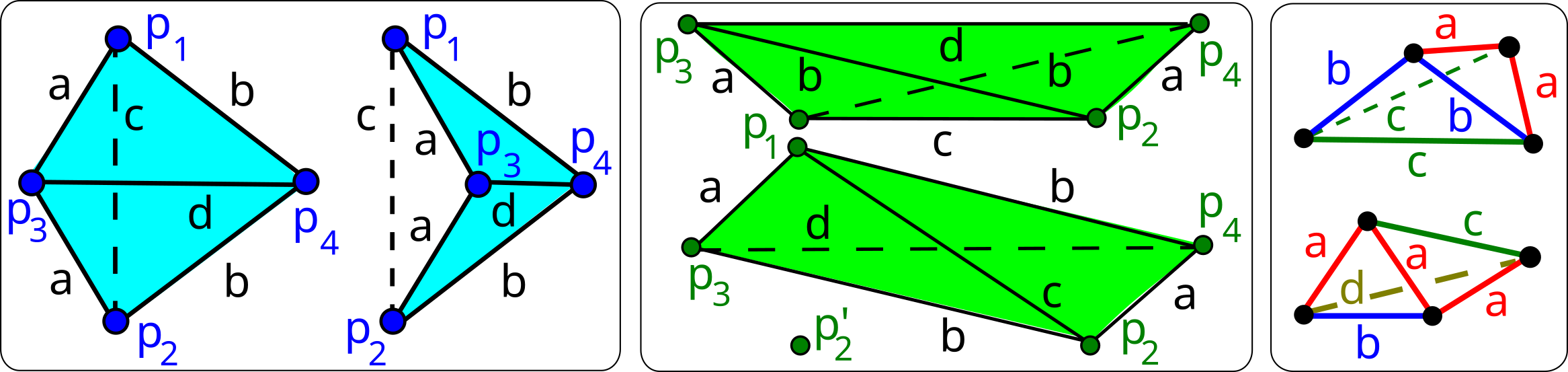}
\caption{\textbf{Left}: 
in $\R^2$, the convex and non-convex kites have two equal rows $\{a,b,c\}$ in $\PDD(C;3)$ and are distinguished by $d=|p_3-p_4|$, see Lemma~\ref{lem:PDD_symmetry}.
\textbf{Middle}: an isosceles trapezoid and parallelogram have $\PDD(C;3)$ with two pairs of equal rows $\{a,b,c\}$ and $\{a,b,d\}$, e.g. a rectangle for $c=d$. 
\textbf{Top right}: a \emph{trisosceles} cloud.
\textbf{Bottom right}: a \emph{3-chain-equal} cloud, see Example~\ref{exa:trisosceles+trichainequal}. 
}
\label{fig:symmetric+special_clouds}
\end{figure}

\begin{lem}[$\PDD$ detects symmetry of $m=4$ points]
\label{lem:PDD_symmetry}
For any cloud $C\subset\R^n$ of $m=4$ distinct points for $n=2,3$, if $\PDD(C;3)$ has two equal rows, then $C$ is either (1) mirror symmetric in the plane passing through two points of $C$ orthogonally to the line segment joining the other points of $C$, or (2) symmetric by the $180^\circ$ degree rotation around the line through the mid-points of two pairs of points of $C$.   
If $n=2$, then $C$ defines a kite, or a parallelogram or an isosceles trapezoid; see Fig.~\ref{fig:symmetric+special_clouds}. 
\end{lem}

\begin{proof}
Let points $p_1,p_2\in C$ have the same row $a\leq b\leq c$ in $\PDD(C;3)$.
One of the distances $a,b,c$ is between the points $p_1,p_2$.
Without loss of generality, assume that $|p_1-p_2|=c$.
Then $p_1,p_2$ have distances $a,b$ to the points $p_3,p_4\in C\setminus\{p_1,p_2\}$. 
\smallskip

\emph{Isosceles case}.
Let $|p_1 - p_3| = a = |p_2 - p_3|$ and $|p_1 - p_4| = b = |p_2 - p_4|$, see Fig.~\ref{fig:symmetric+special_clouds}~(left).
Then $C$ has two equal triangles $\triangle p_1 p_3 p_4=\triangle p_2 p_3 p_4$
 and two isosceles triangles $\triangle p_3 p_1 p_2$ and $\triangle p_4 p_1 p_2$ with equal sides at $p_3,p_4$, respectively.
Let $L$ be the plane that passes through $p_3,p_4$ and is orthogonal to the line segment $[p_1,p_2]$.
Then the mirror reflection in $L$ swaps $p_1,p_2$. 
If $n=2$, $C$ defines a (non-)convex kite. 
\smallskip
 
\emph{Non-isosceles case}.
Then $|p_1-p_3|=a=|p_2-p_4|$ and $|p_2-p_3|=b=|p_1-p_4|$, see Fig.~\ref{fig:symmetric+special_clouds}~(middle).
Let $L$ be the perpendicular bisector of the line segment $[p_3,p_4]$.
The mirror reflection in $L$ swaps $p_3\lra p_4$ and either swaps $p_1\lra p_2$ 
(then $C$ defines an isosceles trapezoid in $\R^2$) or maps $p_2$ to $p'_2$, so that $p_1,p'_2,p_3,p_4$ satisfy the previous case.
In the latter case, the composition with the reflection in the plane through $p_3,p_4$ orthogonal to $[p_1,p'_2]$ is the $180^\circ$ degree rotation that
swaps the points as $p_1\lra p_2$ and $p_3\lra p_4$.
If $n=2$, then $C$ defines a parallelogram, see  Fig.~\ref{fig:symmetric+special_clouds}~(bottom middle).
\end{proof}

\begin{exa}[trisosceles and 3-chain-equal clouds in $\R^3$]
\label{exa:trisosceles+trichainequal}
Fig.~\ref{fig:symmetric+special_clouds}~(right) shows \emph{trisosceles} and \emph{3-chain-equal} clouds that have 3 pairs of equal distances and a chain of 3 equal distances, their $\PDD$s are
$\left(\begin{array}{ccc} a & a & c \\ a & b & b \\ a & b & c \\ b & c & c \end{array} \right)$,
$\left(\begin{array}{ccc} a & a & b \\ a & a & c \\ a & b & d \\ a & c & d \end{array} \right)$, 
respectively.
\end{exa}

\begin{proof}[Proof of Theorem~5.3]
\emph{Case $m=2$}.
Any cloud $C\subset\R^n$ of $m=2$ unordered points $p_1,p_2$ (labelled only for convenience) has $\PDD(C;1)$ consisting of the single distance $|p_1-p_2|$, which uniquely determines $C$ under isometry in any $\R^n$.  
\smallskip

\emph{Case $m=3$}.
Any cloud $C\subset\R^n$ of $m=3$ unordered points with pairwise distances $a\leq b\leq c$ has $\PDD(C;2)
=\left(\begin{array}{cc} a & b \\ a & c \\ b & c \end{array} \right)$.
The (lexicographically) first row of $\PDD(C;2)$ gives us $a\leq b$.
Each of the remaining two rows of $\PDD(C;2)$ should contain at least one value of $a$ or $b$, also in all degenerate cases such as $a=b$.
Removing these repeated values from the other two rows gives us $c$, also in the case $b=c$. 
So $\PDD(C;2)$ identifies $a\leq b\leq c$ and hence $C$, uniquely under isometry in any $\R^n$.  
\medskip

\emph{Case $m=4$}, then $n\leq 3$. 
For a cloud $C\subset\R^3$ of $m=4$ unordered points, $\PDD(C;3)$ is a $4\times 3$ matrix.
Assume that $\PDD(C;3)$ has two equal rows $a\leq b\leq c$. 
\smallskip

\emph{Isosceles case}.
In the first case of Lemma~\ref{lem:PDD_symmetry} in Fig.~\ref{fig:symmetric+special_clouds}~(left), 
$\PDD(C;3)$ has two more rows $\{a,a,d\}$ and $\{b,b,d\}$ including two repeated distances (say, $a,b$) among $a,b,c$.
We can form two isosceles triangles with sides $a,a,c$ and $b,b,c$, which can be rotated in $\R^3$ around their common side of the length $c$, but their positions are fixed under isometry in $\R^3$ by the distance $d$ between their non-shared vertices.
\smallskip

\emph{Non-isosceles case}.
In the second case of Lemma~\ref{lem:PDD_symmetry} in Fig.~\ref{fig:symmetric+special_clouds}~(middle), the matrix $\PDD(C;3)$ has two pairs of equal rows of distances $\{a,b,c\}$ and $\{a,b,d\}$.
Each of these triples uniquely determines a pair of equal triangles with a common side that are symmetric in the perpendicular bisector to this side.
For example, if we start with a fixed position of $[p_3,p_4]$ in $\R^3$, the union of equal triangles $\triangle p_1 p_3 p_4=\triangle p_2 p_3 p_4$ in Fig.~\ref{fig:symmetric+special_clouds}~(middle) is uniquely determined under isometry by the length $d$ of $[p_1,p_2]$.   
In $\R^2$, the parallelogram and isosceles trapezoid are distinguished by 
$d$. 
\medskip

Now we can assume that all rows of $\PDD(C;3)$ are different.
Then all points can be uniquely labelled as $p_1,p_2,p_3,p_4$ according to the lexicographic order of rows. 
Our aim is to get $\PDD(\{p_2,p_3,p_4\};2)$, reconstruct $\triangle p_2 p_3 p_4$, and then uniquely add $p_1$.
\smallskip

\emph{Case of a row with 3 equal distances}.
Let $\PDD(C;3)$ have a row of (say) $p_1$ with 3 equal distances $a$.
After removing the row of $p_1$, the distance $a$ from the rows of $p_2,p_3,p_4$, we get $\PDD(\{p_2,p_3,p_4\};2)$.
This smaller $3\times 2$ matrix determines $\triangle p_2 p_3 p_4$, uniquely under isometry in $\R^3$.
For a fixed $\triangle p_2 p_3 p_4$, the position of $p_1$ in $\R^3$ is determined by its distance $a$ to $p_2,p_3,p_4$, uniquely under the mirror reflection relative to the plane of $\triangle p_2 p_3 p_4$.   
If $n=2$, then $p_1$ is the unique circumcenter of $\triangle p_2 p_3 p_4$.
\smallskip

\emph{Case of a row with 3 unique distances}.
Let $\PDD(C;3)$ have a row of (say) $p_1$, where each of the distances $a,b,c$ (say, to $p_2$, $p_3$, $p_4$) appears in at most one other row (then $a,b,c$ are distinct).
After removing the row of $p_1$, the distance $a$ from the row $p_2$, the distance $b$ from the row of $p_3$, and the distance $c$ from the row of $p_4$, we get $\PDD(\{p_2,p_3,p_4\};2)$.
This $3\times 2$ matrix determines $\triangle p_2 p_3 p_4$, uniquely under isometry in $\R^3$.
Then the position of $p_1$ in $\R^3$ is determined by its distances $a,b,c$ to $p_2,p_3,p_4$, respectively, under a mirror reflection relative to the plane of the triangle $\triangle p_2 p_3 p_4$.
\smallskip

\emph{Case of one distance in 4 rows}.
Then two pairs of points have disjoint edges of the same length, e.g. $|p_1-p_2|=a=|p_3-p_4|$, so 
$\PDD(C;3)=\left(\begin{array}{ccc} a & b & c \\ a & d & e \\ a & b & d \\ a & c & e \end{array} \right)$ for $b=|p_1-p_3|$, $c=|p_1-p_4|$, $d=|p_2-p_3|$, $e=|p_2-p_4|$.
Then $c\neq d$ and $b\neq e$, else $\PDD(C;3)$ has two equal rows (considered above), similarly when $b=c$ and $d=e$.
If $a$ equals one of $b,c,d,e$ (say, $e$), $C$ is a 3-chain-equal cloud in Fig.~\ref{fig:symmetric+special_clouds}~(bottom right) and the argument below still works.
If $b\neq c$, we remove the row of $p_1$, the distance $b$ from the only row of $p_3$ containing $b$, the distance $c$ from the only row of $p_4$ containing $c$, and then remove $a$ from the remaining row of $p_2$.
This reduction to $\PDD(\{p_2,p_3,p_4\};2)$ allows us to reconstruct $C$, uniquely under isometry in $\R^3$, as in the case of a row with 3 unique distances.
If $b=c$ but $d\neq e$, we remove the row of $p_2$, the distance $d$ from the only row of $p_3$ containing $d$, the distance $e$ from the only row of $p_4$ containing $e$, and then remove $a$ from the remaining row of $p_1$, which allows us to uniquely reconstruct $C$, as in the case of a row with 3 unique distances above.
\smallskip

\emph{The final case}: no distance appears in all 4 distinct rows and but every row has a distance appearing in 3 rows, hence at least four times, including two times in the same row.
Then $C$ is a trisosceles cloud in Fig.~\ref{fig:symmetric+special_clouds}~(top right).
If any of the remaining distances $a,b,c$ are equal, $\PDD(C;3)$ has two equal rows (the case considered above).
Then we remove any row (say $a,b,b$) with two repeated distances, the distance $b$ from the only two rows containing $b$, and the distance $a$ from the remaining row.
This reduction to $\PDD(\{p_2,p_3,p_4\};2)$, allows us to reconstruct $\triangle p_2 p_3 p_4$, uniquely under isometry in $\R^3$.
Though $p_1$ has equal distances to two of the vertices (say $p_2,p_3$), the ambiguity of reconstructing $p_1$ in $\R^3$ by its distances to $p_2,p_3,p_4$, is only under the mirror reflections relative to the bisector plane of $[p_2,p_3]$ and the plane of $\triangle p_2 p_3 p_4$.
\end{proof}

\end{document}

%% file: commands.txt
\usepackage{utfsym}
\usepackage{amsmath,amssymb}
\usepackage{cleveref}

\theoremstyle{definition}
\newtheorem{thm}{Theorem}[section]
\newtheorem{dfn}[thm]{Definition}
\newtheorem{pro}[thm]{Problem}
\newtheorem{cor}[thm]{Corollary}
\newtheorem{lem}[thm]{Lemma}
\newtheorem{exa}[thm]{Example}

\newtheorem{conj}[thm]{Conjecture}

\newcommand{\R}{\mathbb{R}}
\newcommand{\Z}{\mathbb{Z}}

\newcommand{\GL}{\mathrm{GL}}

\newcommand{\Eu}{\mathrm{E}}
\newcommand{\SE}{\mathrm{SE}}

\newcommand{\SPD}{\mathrm{SPD}}
\newcommand{\PDD}{\mathrm{PDD}}

\newcommand{\AMD}{\mathrm{AMD}}
\newcommand{\ADA}{\mathrm{ADA}}
\newcommand{\PDA}{\mathrm{PDA}}
\newcommand{\AND}{\mathrm{AND}}
\newcommand{\PND}{\mathrm{PND}}

\newcommand{\EMD}{\mathrm{EMD}}

\newcommand{\PPC}{\mathrm{PPC}}

\newcommand{\Ga}{\Gamma}
\newcommand{\de}{\delta}

\newcommand{\ep}{\varepsilon}

\newcommand{\al}{\alpha}
\newcommand{\la}{\lambda}
\newcommand{\La}{\Lambda}

\newcommand{\si}{\sigma}

\newcommand{\ti}{\tilde}

\newcommand{\bd}{\partial}

\newcommand{\vol}{\mathrm{vol}}

\newcommand{\lra}{\leftrightarrow}

\newcommand{\angstrom}{\textup{\AA}}